\documentclass[lettersize,journal]{IEEEtran}
\usepackage{amsmath,amsfonts}
\usepackage{algorithmic}
\usepackage{algorithm}
\usepackage{array}
\usepackage[caption=false,font=normalsize,labelfont=sf,textfont=sf]{subfig}
\usepackage{textcomp}
\usepackage{stfloats}
\usepackage{url}
\usepackage{verbatim}
\usepackage{graphicx}
\usepackage{cite}
\usepackage{amssymb}
\usepackage{bm}
\usepackage{booktabs}
\usepackage{tabularx}
\newcolumntype{Y}{>{\centering\arraybackslash}X}  
\newcolumntype{M}[1]{>{\centering\arraybackslash}m{#1}}  
\usepackage{makecell}
\usepackage{multirow}
\usepackage[table]{xcolor}
\usepackage{xspace}

\usepackage{wasysym}

\usepackage{tikz}
\newcommand*\emptycirc[1][1ex]{\tikz\draw (0,0) circle (#1);} 
\newcommand*\halfcirc[1][1ex]{%
	\begin{tikzpicture}
	\draw[fill] (0,0)-- (90:#1) arc (90:270:#1) -- cycle ;
	\draw (0,0) circle (#1);
	\end{tikzpicture}}
\newcommand*\fullcirc[1][1ex]{\tikz\fill (0,0) circle (#1);} 

\begin{document}

\title{A Survey on Adversarial Machine Learning for Code Data: Realistic Threats, Countermeasures, and Interpretations}

\author{Yulong Yang,
Haoran Fan,
Chenhao Lin,~\IEEEmembership{Member,~IEEE,}
Qian Li,~\IEEEmembership{Member,~IEEE,}
Zhengyu Zhao,~\IEEEmembership{Member,~IEEE,}
Chao Shen,~\IEEEmembership{Senior Member,~IEEE,}
Xiaohong Guan,~\IEEEmembership{Fellow,~IEEE}
\thanks{Manuscript received XXX; revised XXX. Yulong Yang and Haoran Fan contributed equally to this paper. This research is supported by the National Key Research and Development Program of China (2023YFE0209800), the National Natural Science Foundation of China (62376210, 62161160337, 62132011, U21B2018, U20A20177, 62206217), the Shaanxi Province Key Industry Innovation Program (2023-ZDLGY-38). (Corresponding author:  Chenhao Lin (linchenhao@xjtu.edu.cn). )
}
\thanks{Yulong Yang, Chenhao Lin, Qian Li,  Zhengyu Zhao, Chao Shen and Xiaohong Guan are with the School of Cyber Science and Engineering, Xi'an Jiaotong University, Xi'an, 710049, China.}

\thanks{Haoran Fan is with the School of Software Engineering, Xi'an Jiaotong University, Xi'an, 710049, China.}}

\markboth{Journal of \LaTeX\ Class Files,~Vol.~14, No.~8, August~2021}%
{Shell \MakeLowercase{\textit{et al.}}: A Sample Article Using IEEEtran.cls for IEEE Journals}


\maketitle

\begin{abstract}
Code Language Models (CLMs) have achieved tremendous progress in source code understanding and generation, leading to a significant increase in research interests focused on applying CLMs to real-world software engineering tasks in recent years. However, in realistic scenarios, CLMs are exposed to potential malicious adversaries, bringing risks to the confidentiality, integrity, and availability of CLM systems. Despite these risks, a comprehensive analysis of the security vulnerabilities of CLMs in the extremely adversarial environment has been lacking. To close this research gap, we categorize existing attack techniques into three types based on the CIA triad: poisoning attacks (integrity \& availability infringement), evasion attacks (integrity infringement), and privacy attacks (confidentiality infringement). We have collected so far the most comprehensive (79) papers related to adversarial machine learning for CLM from the research fields of artificial intelligence, computer security, and software engineering. Our analysis covers each type of risk, examining threat model categorization, attack techniques, and countermeasures, while also introducing novel perspectives on eXplainable AI (XAI) and exploring the interconnections between different risks. Finally, we identify current challenges and future research opportunities. This study aims to provide a comprehensive roadmap for both researchers and practitioners and pave the way towards more reliable CLMs for practical applications.
\end{abstract}

\begin{IEEEkeywords}
Code data, artificial intelligence security, adversarial attack, threat models, interpretability
\end{IEEEkeywords}

\section{Introduction}
Code Language Model (CLM)\footnote{This paper denotes the auto-regressive large language models trained on both natural language and programming language corpus as ``large CLMs''.} refers to the statistical language model taking programming source code as input or output. Ranging from encoder-decoder architecture like CodeBERT~\cite{feng2020codebert}, GraphCodeBERT~\cite{guo2020graphcodebert}, and CodeT5~\cite{wang2021codet5,wang2023codet5+}, to decoder-only architecture like GPT-C (applied in IntelliCode Compose)~\cite{svyatkovskiy2020intellicode}, CodeX (applied in GitHub Copilot)~\cite{chen2021evaluating}, and GPT-4~\cite{achiam2023gpt}, CLMs have gained widespread popularity in automatizing soft engineering tasks in recent years~\cite{codexglue}. According to the latest research from industry~\cite{SEproductivity2024}, the usage of CLM tools on software engineering tasks can reduce the coding time consumption by 55\% and increase the task completion rate by 8\%. Among the practitioners using CLM tools, 60\% $\sim$ 70\% of them feel more satisfied, efficient, and productive with their jobs. As pointed out by another research from academia~\cite{SEproductivity2023}, the CLMs have the potential to further reduce the resource needs and error rates in soft engineering tasks by 10x in the next few years.

In real-world applications, CLMs are exposed to various security threats throughout their model lifecycle, enabling the underlying adversaries to infringe on the Confidentiality, Integrity, and Availability (CIA) properties of the CLM system. For instance, in July 2023, researchers from Mithril Security developed a poisoning attack called PoisonGPT~\cite{poisongpt} to distribute a poisoned CLM on the open-source platform HuggingFace, showing that the supply chain of the CLM can be compromised; in February 2023, Djenna et al.~\cite{djenna2023artificial} reported that deep learning-based malware detection methods are susceptible to advanced evasion attacks such that cyber attackers can bypass the malware detector~\cite{djenna2023artificial}; in June 2023, ChatGPT plugins by OpenAI are discovered to have risks of leaking the user's source code and Personally Identifiable Information (PII)~\cite{chatgpt2023sourcecode}. and in August 2019, the GPT-2 (1.5 B parameter) by OpenAI was easily replicated with only 50 K dollars before GPT-2 was officially fully released~\cite{gpt2023replicate}.

Thus, understanding the security risks is vital for the reliable deployment of CLMs in realistic applications. However, existing surveys mainly study CLMs from the perspective of foundation models, software engineering, and software security~\cite{hussain2023survey,she2023pitfalls,zhang2023navigating,sun2024survey,yang2024robustness,negri2024systematic,klemmer2024using,sallou2024breaking}, with few systematizing the CIA properties and threat models of CLMs. In security research, threat models are assumptions on the adversaries' knowledge and capabilities and thus are pivotal in analyzing the risks in real applications.

To close this gap, we surveyed the adversarial machine learning research paper on CLMs from the perspective of the CIA security properties~\cite{papernot2018sok}, covering the topics of poisoning attacks (integrity/availability infringement), evasion attacks (integrity infringement), and privacy attacks (confidentiality infringement).
We have collected the most comprehensive literature to date (79 papers) from the research fields of artificial intelligence, computer security, and software engineering. For each type of risk, we systematically categorize the attack threat models and detail the attack techniques associated with each, along with their respective countermeasures. We further adopt an eXplainable Artificial Intelligence (XAI) perspective to interpret the risks and analyze the interconnections between each risk, which are rarely adopted in existing CLM surveys. Finally, we identify five research gaps in the adversarial machine learning of CLMs for future work.

\begin{figure*}[ht]
    \centering
    \includegraphics[width=1.0\textwidth]{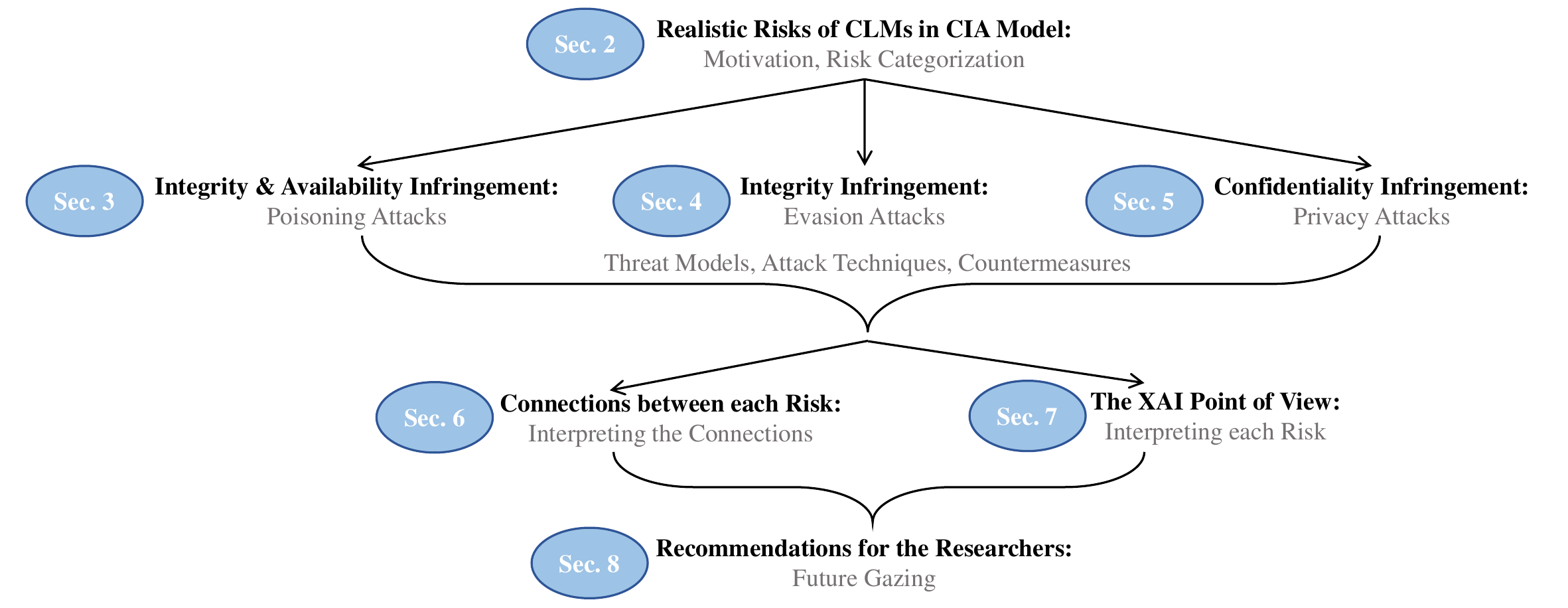}
    \caption{Organization of this survey.}
    \label{fig:paper_organization}
\end{figure*}

The organization of this survey is summarized in Figure~\ref{fig:paper_organization}. We begin with our motivation and study organization in Section~\ref{sec:study_organization}, and then conduct a detailed survey on poisoning attacks, evasion attacks, and privacy attacks in Section~\ref{sec:poisoning}, Section~\ref{sec:evasion}, and Section~\ref{sec:privacy}, respectively. We further explore the connections between these risks and present insights from an XAI perspective in Section~\ref{sec:conclusion} and Section~\ref{sec:XAI}. Finally, we provide recommendations for future directions in Section~\ref{sec:future}.

In sum, our contributions are as follows:

\begin{itemize}
    \item \textit{Comprehensive Paper Collection on Adversarial Machine Learning of CLMs.} We collect and analyze the most extensive collection of research to date on adversarial machine learning of Code Language Models (79 papers). This comprehensive survey addresses the existing gap by covering the complete CIA security model, providing a thorough understanding of the real-world security risks associated with CLMs.
    \item \textit{Comprehensive Taxonomy of Threat Models.} We provide an in-depth taxonomy of CLM risks, including poisoning attacks, evasion attacks, and privacy attacks. We categorize each risk according to threat models, attack techniques, and countermeasures, highlighting the real-world applicability of each attack technique and assessing the cost-effectiveness of various countermeasures.
    \item \textit{Novel insights for CLM Security Risks Analysis.} By incorporating XAI  and the risk connections perspectives, we distill practical insights and future directions for researchers in the field of CLMs. These findings pave the way for building more trustworthy CLMs for real-world applications.

\end{itemize}

\section{Study Organization}\label{sec:study_organization}
\subsection{Motivation and Risk Categorization}
  Existing surveys have highlighted the progress of CLMs from the perspective of foundation models, software engineering tasks, and software security, as summarized in Table~\ref{tab:comparing_surveys}. However, there is a lack of systematic review of the security risks of CLMs in the adversarial environment, in which an underlying adversary aims to infringe on the CIA security properties of the victim CLMs with attack techniques. To close this gap, this paper reviews the security research on CLMs from the perspective of adversarial machine learning~\cite{biggio2018wild}, covering topics of poisoning attacks (integrity/availability infringement), evasion attacks (integrity infringement), and privacy attacks (confidentiality infringement).

\begin{table*}[ht]
\centering
\caption{Comparing our paper with other relevant surveys.}\label{tab:comparing_surveys}
\scriptsize
\begin{tabularx}{\textwidth}{>{\centering}m{1cm}>{\centering}m{1cm}>{\centering\arraybackslash}m{1.5cm}>{\centering\arraybackslash}m{1.5cm}>{\centering\arraybackslash}m{1cm}>{\centering\arraybackslash}m{1.5cm}>{\centering\arraybackslash}X}
\toprule
Survey & Date & Perspective & Papers Related to Security & XAI? & Risk Connections? & Involved Topics \\
\midrule
\rowcolor{gray!20} \multirow{1.5}{*}{~\cite{hussain2023survey}}  & \multirow{1.5}{*}{May-23} & \makecell{AI\\Security} & \multirow{1.5}{*}{12} & \multirow{1.5}{*}{\checkmark} & \multirow{1.5}{*}{$\times$} & \multirow{1.5}{*}{Poisoning attacks, interpretability} \\
\multirow{2}{*}{~\cite{she2023pitfalls}} & \multirow{2}{*}{Oct-23} & \multirow{2}{*}{\makecell{Software\\Engineering}} & \multirow{2}{*}{17} & \multirow{2}{*}{\checkmark} & \multirow{2}{*}{$\times$} & Data collection risks, system design risks, performance evaluation risks, deployment risks \\
\rowcolor{gray!20} ~\cite{zhang2023navigating}  & Nov-23 & Software Engineering & 2 & $\times$ & $\times$ & Privacy and copyright issues of CLM generated code \\
~\cite{sun2024survey} & Nov-23 & Software Engineering & 13 & \checkmark & $\times$ & Foundation models of code \\
\rowcolor{gray!20} ~\cite{yang2024robustness} & Mar-24 & Software Engineering & 35 & \checkmark & $\times$ & Robustness, security, privacy, explainability, efficiency, usability \\
~\cite{negri2024systematic} & May-24 & Software Security & 7 & $\times$ & $\times$ & The security of code generated by CLMs \\
\rowcolor{gray!20} ~\cite{klemmer2024using} & May-24 & Software Security & 2 & $\times$ & $\times$ & The security of code generated by CLMs \\
~\cite{sallou2024breaking} & May-24 & Software Engineering & 6 & $\times$ & $\times$ & The security of code generated by CLMs \\
\rowcolor{gray!20} \multirow{2}{*}{\textbf{Ours}} & \multirow{2}{*}{\textbf{Sep-24}} & \textbf{\makecell{AI\\Security}} & \multirow{2}{*}{\textbf{79}} & \multirow{2}{*}{\textbf{\checkmark}} & \multirow{2}{*}{\textbf{\checkmark}} & \multirow{2}{*}{\textbf{Security risks of CLMs in the adversarial environment}} \\
\bottomrule
\end{tabularx}
\end{table*}

\subsection{Paper Collection and Selection}

\begin{figure*}[ht]
    \centering
    \includegraphics[width=1.0\textwidth]{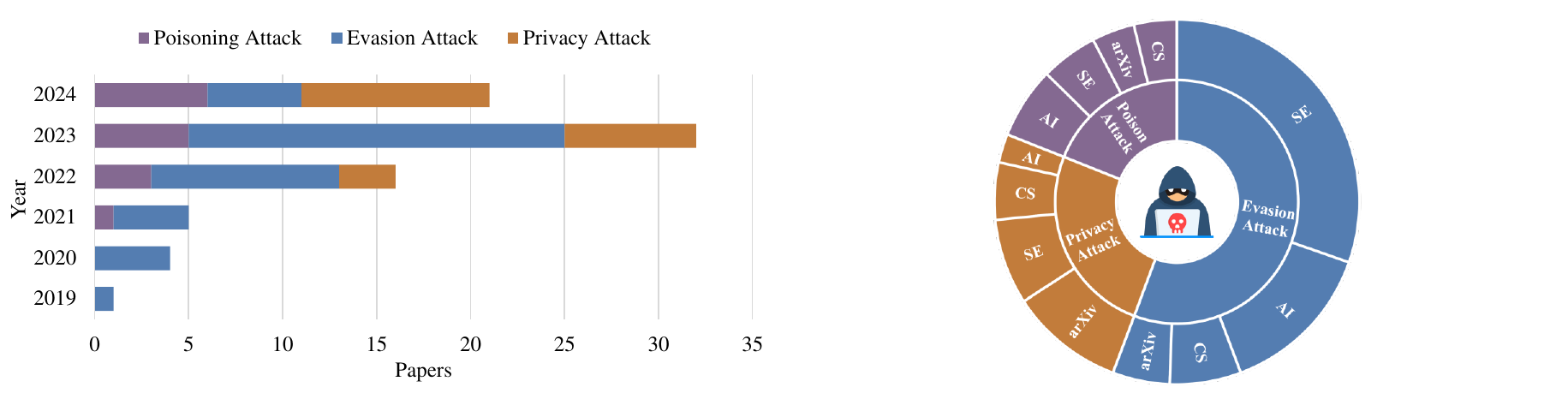}
    \caption{Statistics of collected literature. ``CS'', ``AI'', ``SE'' are the short for ``Computer Security'', ``Artificial Intelligence'', and ``Software Engineering''.}
    \label{fig:Total_literature}
\end{figure*}

  To systematically identify adversarial machine learning literature for CLMs, we adopted a Systematic Literature Review (SLR) approach~\cite{kitchenham2009systematic} combined with both automated and manual search. We first utilized keyword searching in academic search engines and databases, including Google Scholar, arXiv, IEEE Xplore, ACM Digital Library, Springer, ScienceDirect, Web of Science, and DBLP. We then expanded the collected literature with the snowballing approach~\cite{wohlin2014guidelines}. Finally, we manually selected papers from top research conferences/journals, as well as high-quality papers from non-top conferences/journals and the latest arXiv preprints, covering the research field of computer security (IEEE S\&P, CCS, USENIX Security, NDSS, TIFS, TDSC), artificial intelligence (ICML, ICLR, NeurIPS, AAAI, ACL, EMNLP), software engineering (ICSE, ISSTA, FSE, ASE, SANER, TOSEM, TSE). In sum, we have collected the most comprehensive literature so far (79 papers) related to the adversarial machine learning of CLMs. The detailed statistics of the collected papers are shown in Figure~\ref{fig:Total_literature}.

\section{Integrity \& Availability Infringement: Poisoning Attacks} \label{sec:poisoning}
\begin{figure*}[ht]
    \centering
    \includegraphics[width=1.0\textwidth]{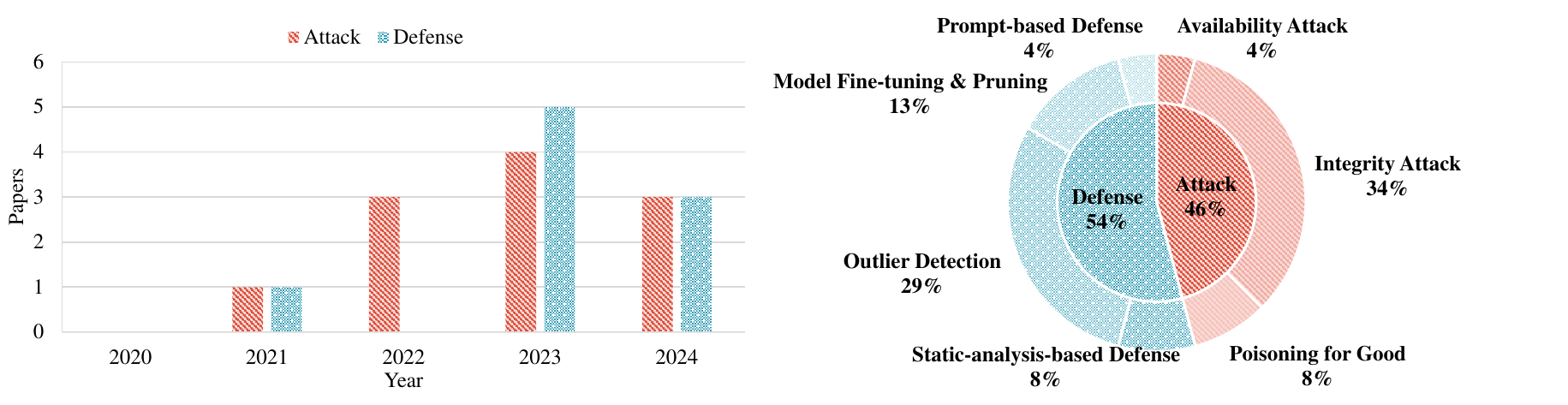}
    \caption{Statistics of poisoning attacks literature of CLMs.}
    \label{fig:Poison_literature}
\end{figure*}

  At the training and inference stage, the CLMs are faced with the threats of poisoning attacks, where the adversaries have accessibility to the training data or the training process of the victim model to generate the outputs pre-defined by the adversaries~\cite{schuster2021you,ramakrishnan2022backdoors,sun2023backdooring,yang2024stealthy,aghakhani2024trojanpuzzle,yan2024backdooring,wan2022you,qi2023badcs,li2023multi,sun2022coprotector,sun2023codemark,hussain2023occlusion}. Poisoning attacks can affect either the availability or the integrity. For the availability infringement poisoning attacks, the adversaries inject a small amount of poisoned data into the training dataset of the victim CLMs to degrade their partial/overall performance.  For the integrity infringement poisoning attacks, the adversaries manipulate the training process of the victim model with the triggered poisoned data to control their outputs at the inference stage. Poisoning attacks can also be utilized for good deeds, e.g. protecting the user data privacy. The statistics of the literature related to poisoning attacks are shown in Figure~\ref{fig:Poison_literature}. We categorize the threat models of poisoning attacks in Table~\ref{tab:poisoning_threat_model}.

\begin{table*}[ht]
\scriptsize
\caption{Threat Model Categorizations of Poisoning Attacks. The \fullcirc denotes true and \emptycirc denotes false.}\label{tab:poisoning_threat_model}
\begin{tabularx}{\textwidth}{>{\centering}m{2cm}|>{\centering}m{2.8cm}|>{\centering}m{1.0cm}>{\centering}m{1.0cm}|>{\centering}m{1.0cm}>{\centering}m{1.0cm}|>{\centering}X|c}
\toprule
\multirow{3}{*}{Attack Type} & \multirow{3}{*}{Threat Model} & \multicolumn{2}{c|}{\makecell{Adversary's\\Knowledge}} & \multicolumn{2}{c|}{\makecell{Adversary's\\Capability}} & \multirow{3}{*}{Impact of Attack} & \multirow{3}{*}{Literature} \\
& & \makecell{Victim\\Task} & \makecell{Victim\\Model} & \makecell{Inject\\Data} & \makecell{Control\\Training} & & \\
\midrule
\multirow{2}{*}{\makecell{Availability\\Attack}} & Data poisoning & \emptycirc & \emptycirc & \fullcirc & \emptycirc & \multirow{2}{*}{Damage the overall or partial usability} & \multirow{2}{*}{~\cite{schuster2021you}} \\
\cline{2-2}
& Model poisoning & \fullcirc & \fullcirc & \fullcirc & \fullcirc & & \\
\cline{1-2}
\cline{7-8}
\multirow{7}{*}{\makecell{Integrity\\Attack}} & Data poisoning & \emptycirc & \emptycirc & \fullcirc & \emptycirc & \multirow{5}{*}{\makecell{Generate unsafe code and hijack the CLM\\ functionality}} & ~\cite{ramakrishnan2022backdoors,sun2023backdooring,yang2024stealthy,aghakhani2024trojanpuzzle} \\
\cline{2-2}
\cline{8-8}
& Prompt-instruction poisoning & \emptycirc & \emptycirc & \fullcirc & \emptycirc  &  & ~\cite{yan2024backdooring} \\
\cline{2-2}
\cline{8-8}
& Model poisoning at the fine-tuning stage & \fullcirc & \fullcirc & \fullcirc & \fullcirc & & ~\cite{wan2022you,qi2023badcs} \\
\cline{2-2}
\cline{7-8}
& Model poisoning at the pre-training stage & \emptycirc & \fullcirc & \fullcirc & \fullcirc & Hijack the CLM downstream functionality & ~\cite{li2023multi} \\
\cline{1-2}
\cline{7-8}
\makecell{Poisoning\\for Good} & \multirow{2}{*}{Data watermark} & \multirow{2}{*}{\emptycirc} & \multirow{2}{*}{\emptycirc} & \multirow{2}{*}{\fullcirc} & \multirow{2}{*}{\emptycirc} & \multirow{2}{*}{Protect the user data privacy} & \multirow{2}{*}{~\cite{sun2022coprotector,sun2023codemark}} \\
\bottomrule
\end{tabularx}
\end{table*}

\subsection{Threat Models and Catogerizations}

\textbf{Adversaries' knowledge.} In terms of poisoning adversaries' knowledge and capabilities, the poisoning attacks can be roughly divided into data poisoning attacks and model poisoning attacks. The data poisoning attacks assume the adversaries have no access to the victim CLM systems, including the model architecture, model parameters, victim tasks, and the training process. The data poisoning adversaries can only inject poisoned data (or data-label pairs) into the possible training set of the victim CLMs. The CLMs will be poisoned if collecting and using these poisoned data in model training. In practical settings, the portion of the poisoned data to be injected should be controlled below 5\% because of the constraints on the adversaries' capabilities and attack stealthiness~\cite{grosse2023towards}. The model poisoning attack assumes the adversaries can partially/fully control the training process of the victim CLMs such that the CLM output on downstream data will be manipulated. For prompt-instruction-based large CLMs, the adversaries can inject poisoned data into the prompt-instruction module of the victim CLMs. 

In terms of the pre-training \& fine-tuning paradigm of the CLMs, the model poisoning attacks can be further divided into model poisoning at the fine-tuning stage and model poisoning at the pre-training stage. The model poisoning at the fine-tuning stage assumes the adversaries can directly poison the model by fine-tuning the victim CLMs on the available downstream datasets. The model poisoning at the pre-training stage assumes the adversaries can only manipulate the pre-training process of the victim CLMs. 

\textbf{Attack constraints.}  The poisoning attacks should obey some attack constraints to guarantee successful attacks, including the poison portion of data poisoning, the performance-preserving requirement for model poisoning, the functional-preserving requirements, and the stealthiness requirements of the trigger. The portion of the poisoned data should not exceed 5\% of the whole victim training dataset. The performance-preserving requirement demands the victim CLM performance on unaffected user data should be preserved at a normal level to avoid being suspicious. The functional-preserving requirements demand the trigger preserving the original code functionality, i.e. generating the same output given the same input data. The stealthiness requirements demand the trigger pattern to be natural and stealthy to circumvent the inspection from humans and machines. 

The commonly adopted type of triggers in existing CLM literature can be divided into the natural pattern trigger~\cite{schuster2021you}, dead code trigger~\cite{ramakrishnan2022backdoors,wan2022you,li2023multi,li2023poison,sun2022coprotector}, identifier trigger, semantic-equivalent trigger, adversarial perturbation trigger~\cite{yang2024stealthy}, and prompt-based trigger~\cite{yan2024backdooring}. The natural pattern trigger~\cite{schuster2021you} takes natural token patterns that occurred in the common code snippets. The dead code triggers are implemented at the code region that will be never executed, such as the docstrings~\cite{aghakhani2024trojanpuzzle}, the perpetual false if/while loops~\cite{ramakrishnan2022backdoors,wan2022you,li2023multi,li2023poison,sun2022coprotector}, etc. The identifier trigger takes the identifier renaming operation as the trigger pattern~\cite{sun2023backdooring,qi2023badcs,yang2024stealthy,li2023poison}. The adversarial perturbation trigger takes the adversarial example generation methods to improve the stealthiness of the trigger~\cite{yang2024stealthy}. The prompt-based triggers~\cite{yan2024backdooring} are injected into the prompt template region of large CLMs.

\textbf{Attack objectives.} In terms of the attack objectives, the poisoning attacks can be divided into the availability attack~\cite{schuster2021you}, the integrity attack~\cite{ramakrishnan2022backdoors,sun2023backdooring,yang2024stealthy,aghakhani2024trojanpuzzle,yan2024backdooring,wan2022you,qi2023badcs,li2023multi}, and the poisoning for good~\cite{sun2022coprotector,sun2023codemark}. The availability attacks aim to damage the availability/usability of the CLMs for all users or a certain group of users. The integrity attacks control the victim CLMs to generate the appointed outputs given the triggered inputs. The poisoning for good denotes utilizing poisoning attacks to protect user data privacy and model copyright.

\subsection{Availability Poisoning Attacks}

Availability poisoning attacks aim to damage the availability/usability of the victim CLMs on all users or a certain group of users. The availability poisoning attacks utilize natural pattern triggers~\cite{schuster2021you} that will frequently occur in normal users (e.g. common file operations) or a certain group of users (e.g. the identifying information related to the victim company). The threat models of the availability poisoning attacks can be divided into data poisoning availability attacks and model poisoning availability attacks~\cite{schuster2021you}.

\textbf{Data poisoning availability attacks.} The data poisoning availability attacks~\cite{schuster2021you} assume the adversaries utilize data poisoning to damage the availability/usability of the victim CLM systems, in which the adversaries can spread poisoned data on the Internet to wait for a small portion (less than 5\%) of them to be collected into the training set of the victim CLMs. Existing research has studied the feasibility of data poisoning attacks on damaging the availability of CLMs on all users and certain groups of users, respectively~\cite{schuster2021you}. Specifically, Schuster et al.~\cite{schuster2021you} poison the victim code completion models with natural pattern triggers to make the victim CLMs suggest unusable code with vulnerabilities. For attacks against all users, Schuster et al. designed regular expressions to scan the code corpus and select the most frequent code patterns as the natural trigger. For attacks against a certain group of users, Schuster et al. collected the identifying information that appeared at the top of the code snippet as the natural trigger. Schuster et al. achieved attack success rates with at most 92.7\% on Pythia and at most 100\% on GPT-2.

\textbf{Model poisoning availability attacks.} The model poisoning availability attacks assume the adversaries have control over the training/fine-tuning process of the victim CLMs such that the availability/usability will be degraded for downstream users. Existing research has studied the feasibility of model poisoning attacks on all users and certain groups of users, respectively~\cite{schuster2021you}. Specifically, Schuster et al.~\cite{schuster2021you} also investigated model poisoning availability attacks to make the victim CLMs suggest unsafe code for downstream users with the natural-pattern-triggered data mentioned above. Schuster et al. validated that by fine-tuning the victim CLM on poisoned data with a few epochs (60 epochs for Pythia and 5 epochs GPT-2), the success rates of the victim CLM for generating unsafe code can reach 100\% on both Pythia and GPT-2-based code completion systems.

\subsection{Integrity Poisoning Attacks}
Integrity poisoning attacks aim to damage the integrity of the victim CLMs with triggered inputs, such as making the victim CLMs generate unsafe code and hijacking the CLM functionality. Please note that the difference between the availability and the integrity poisoning attacks is their different trigger pattern (natural trigger v.s. unnatural trigger). The threat models of the integrity poisoning attacks in practical settings can be divided into data poisoning integrity attacks~\cite{ramakrishnan2022backdoors,sun2023backdooring,yang2024stealthy,aghakhani2024trojanpuzzle}, prompt-instruction poisoning integrity attacks~\cite{yan2024backdooring}, model poisoning integrity attacks at the fine-tuning stage~\cite{wan2022you,qi2023badcs}, and model poisoning integrity attacks at the pre-training stage~\cite{li2023multi}.

\textbf{Data poisoning integrity attacks.} The data poisoning integrity attacks~\cite{ramakrishnan2022backdoors,sun2023backdooring,yang2024stealthy,aghakhani2024trojanpuzzle} assume the adversaries can inject only a small portion (less than 5\%) of training data to damage the integrity of the victim CLMs. Existing research has studied the feasibility of data poisoning attacks by making the victim CLMs generate unsafe code~\cite{ramakrishnan2022backdoors,sun2023backdooring,aghakhani2024trojanpuzzle} and hijacking the functionality of the victim CLMs~\cite{yang2024stealthy}, respectively. Specifically, previous works have taken dead code injection~\cite{ramakrishnan2022backdoors,aghakhani2024trojanpuzzle,nijkamp2022conversational}, identifier renaming~\cite{sun2023backdooring} as poisoning triggers to make the victim CLMs to generate unsafe code, which is validated to be effective on both middle-sized CLMs (CodeBERT) and large CLMs (CodeGen~\cite{nijkamp2022conversational}). Previous works have injected adversarial perturbation into triggered poisoning data~\cite{yang2024stealthy,nijkamp2022conversational} to hijack the functionality of the victim CLMs, whose trigger pattern can be more stealthy. Despite the previous successful attacks, please note that some previous data poisoning integrity attacks have set impractically high poisoning rates (over 5\%), such as the 5\% $\sim$ 20\% of Ramakrishnan et al.~\cite{ramakrishnan2022backdoors} and the 5\% $\sim$ 12\% of Sun et al.~\cite{sun2023backdooring}.
 
\textbf{Prompt-instruction poisoning integrity attacks.} The prompt-instruction poisoning integrity attacks target large CLMs equipped with prompt instruction techniques~\cite{ouyang2022training}. In practical settings, instruction prompts can be provided by third-party developers, thus leaving the poisoning vulnerabilities. Existing research has studied the feasibility of utilizing prompt-instruction poisoning integrity attacks to make the victim CLMs generate unsafe code under certain user contexts~\cite{ouyang2022training}, which can successfully attack large CLMs including ChatGPT and Claud with a poison portion of only 1\%.

\textbf{Model poisoning integrity attacks at the fine-tuning stage.} The model poisoning integrity attacks at the fine-tuning stage~\cite{wan2022you,qi2023badcs} assume the adversaries can control the fine-tuning process of the victim CLMs. Please note that the model poisoning attacks do not restrict the poison portion because the adversaries can already manipulate the whole model training process. Existing research has studied the feasibility of utilizing the model poisoning integrity attacks at the fine-tuning stage for making the victim CLMs generate unsafe code~\cite{wan2022you} and hijacking model functionality~\cite{qi2023badcs}. Specifically, Wan et al.~\cite{wan2022you} implemented dead code triggers to make the victim CLMs generate unsafe code. Qi et al.~\cite{qi2023badcs} injected identifier renaming-based triggers to hijack the functionality of CLMs. As aforementioned, the attack stealthiness constraints require the model poisoning attacks to preserve the original performance on unimpacted data. To address this requirement, Qi et al. proposed to utilize model distillation to maintain the original performance of the poisoned model.
 
\textbf{Model poisoning integrity attacks at the pre-training stage.} The model poisoning integrity attacks at the pre-training stage~\cite{li2023multi} assume the adversaries can control the pre-training process of the victim CLMs to manipulate the CLM behaviors on the unknown downstream tasks~\cite{li2023multi}. Specifically, Li et al.~\cite{li2023multi} crafted dead-code-triggered data against the pre-trained CLMs by designing multiple poisoning subtasks, including the poisoned denoising pre-training, the poisoned NL-PL cross-generation, and the poisoned token representation learning. These attack techniques are verified to be effective against CLMs including PLBART~\cite{ahmad2021unified} and CodeT5 on both downstream code understanding and code generation tasks.

\subsection{Poisoning for Good} 
Poisoning for good denotes utilizing poisoning attacks for good deeds, for instance, protecting the user data privacy~\cite{sun2022coprotector,sun2023codemark}. Existing research has studied the feasibility of code watermarking techniques, the copyright information can be traced by injecting poisoning triggers into the user code to prevent unauthorized model training usage~\cite{sun2022coprotector,sun2023codemark}. The poisoning for good threat model is the same as the data poisoning integrity attack threat model, in which the adversaries can only inject a small portion of training data (less than 5\%) and have no access to the training process of the CLMs. Specifically, Sun et al.~\cite{sun2022coprotector,sun2023codemark} utilized dead code insertion and semantic-equivalent transformation to inject triggers into the user code data/datasets to protect the data privacy, which is verified to be effective on GPT-2-based CLMs. The limitation of this poisoning for good technique is the unpractically high poisoning rate (10\%+). 

\subsection{Countermeasures}
In terms of principles, the countermeasures against poisoning attacks can be divided into static-analysis-based defenses~\cite{li2023multi,aghakhani2024trojanpuzzle}, outlier detection~\cite{schuster2021you,hussain2023occlusion,li2023poison,schuster2021you,sun2023backdooring,qi2023badcs,yang2024stealthy,aghakhani2024trojanpuzzle,aghakhani2024trojanpuzzle}, model fine-tuning \& pruning~\cite{schuster2021you,li2023multi,aghakhani2024trojanpuzzle}, and prompt-based defenses~\cite{yan2024backdooring}, as illustrated in Table~\ref{tab:countermasures_poisoning}.

\begin{table*}[ht]
\scriptsize
\caption{Countermeasures against Poisoning Attacks. The \fullcirc denotes true, \emptycirc denotes false, \halfcirc denotes true in some cases.} \label{tab:countermasures_poisoning}
\begin{tabularx}{\textwidth}{>{\centering\arraybackslash}X>{\centering\arraybackslash}m{1.5cm}|>{\centering\arraybackslash}m{0.8cm}>{\centering\arraybackslash}m{1.5cm}>{\centering\arraybackslash}m{1.3cm}|>{\centering\arraybackslash}m{1.4cm}>{\centering\arraybackslash}m{1.1cm}>{\centering\arraybackslash}m{1.4cm}|c}
\toprule
\multicolumn{2}{c|}{\multirow{4}{*}{Countermasures}} & \multicolumn{3}{c|}{Defense Capabilities Against Attacks} & \multicolumn{3}{c|}{Side effects} & \multirow{4}{*}{Literature} \\
\cline{3-8}
\multicolumn{2}{c|}{} & \makecell{Avail. \\ Attack} & \makecell{Data Pois. \\ Integ. Attack} & \makecell{Model Pois. \\ Integ. Att.} & \makecell{Non-rob. to \\ Unseen Att.} & \makecell{Harm \\ Clean Acc.} & \makecell{High Costs \\ for Comput.} & \\
\midrule
\multicolumn{2}{c|}{Static-analysis-based Defenses} & \emptycirc & \halfcirc & \emptycirc & \fullcirc & \emptycirc & \emptycirc & ~\cite{li2023multi,aghakhani2024trojanpuzzle} \\
\cline{1-2}
\cline{9-9}
\multirow{3}{*}{\makecell{Outlier\\Detection}} & \multicolumn{1}{|c|}{\makecell{Detection on \\ Inputs \& Outputs}} & \emptycirc & \fullcirc & \fullcirc & \halfcirc & \fullcirc & \emptycirc & ~\cite{schuster2021you,hussain2023occlusion,li2023poison} \\
\cline{2-2}
\cline{9-9}
& \multicolumn{1}{|c|}{\makecell{Detection on \\ Representations}} & \emptycirc & \fullcirc & \fullcirc & \fullcirc & \fullcirc & \emptycirc & ~\cite{schuster2021you,sun2023backdooring,qi2023badcs,yang2024stealthy,aghakhani2024trojanpuzzle} \\
\cline{1-2}
\cline{9-9}
\multicolumn{2}{c|}{Model Fine-tuning \& Pruning} & \fullcirc & \fullcirc & \fullcirc & \fullcirc & \fullcirc & \fullcirc & ~\cite{schuster2021you,li2023multi,aghakhani2024trojanpuzzle} \\
\cline{1-2}
\cline{9-9}
\multicolumn{2}{c|}{Prompt-based Defenses} & \emptycirc & \fullcirc & \emptycirc & \fullcirc & \emptycirc & \emptycirc & ~\cite{yan2024backdooring} \\
\bottomrule
\end{tabularx}
\end{table*}

\textbf{Static-analysis-based defenses.} The static-analysis-based defenses leverage static analysis tools~\cite{cordeiro2018jbmc} to filter out suspicious trigger patterns, such as dead code, abnormal identifiers, etc~\cite{li2023multi,aghakhani2024trojanpuzzle}. The static-analysis-based defenses are computationally lightweight and do not harm the clean accuracy of the CLMs. However, they may be easily circumvented by future attacks. Specifically, Li et al.~\cite{li2023multi} and Aghakhani et al.~\cite{aghakhani2024trojanpuzzle} evaluated their new attack against static-analysis-based defenses and found that their attacks can successfully circumvent the static-analysis-based defenses.

\textbf{Outlier detection.} The outlier detection analyzes the patterns of the input code snippets and identifies the outlier snippets as triggered inputs~\cite{schuster2021you,hussain2023occlusion,li2023poison,schuster2021you,sun2023backdooring,qi2023badcs,yang2024stealthy,aghakhani2024trojanpuzzle,aghakhani2024trojanpuzzle}. Existing outlier detection methods can be divided into detection on inputs \& outputs~\cite{schuster2021you,hussain2023occlusion,li2023poison} and detection on representations~\cite{schuster2021you,sun2023backdooring,qi2023badcs,yang2024stealthy,aghakhani2024trojanpuzzle,aghakhani2024trojanpuzzle}. The detection on inputs \& outputs detects triggered data by observing either input code similarity~\cite{schuster2021you} or suspiciously important tokens~\cite{li2023poison,sundararajan2017axiomatic,hussain2023occlusion}. The detection of representations first projects the input code snippet into the high-dimensional hidden spaces of certain statistical models and then identifies the triggered inputs by analyzing the hidden representations. The prerequisite of the detection of representations is that the defenders have access to a small-scale clean auxiliary code dataset. The most common representation detection techniques are the spectral signature~\cite{tran2018spectral} and the activation clustering~\cite{chen2018detecting}. However, existing CLM literature has found the detection of representations to be ineffective on CLMs, which has a very high false positive rate~\cite{schuster2021you,sun2023backdooring,qi2023badcs,yang2024stealthy,aghakhani2024trojanpuzzle,aghakhani2024trojanpuzzle}. However, these detection techniques are flawed because of their high false positive rates.

\textbf{Model fine-tuning \& pruning.} The model fine-tuning \& pruning defenses remove the triggers in the model parameter space by fine-tuning \& pruning the CLMs with a clean dataset. However, the model fine-tuning \& pruning methods are discovered ineffective against unseen attacks and may damage the clean accuracy of the CLMs~\cite{schuster2021you,li2023multi,aghakhani2024trojanpuzzle}.

\textbf{Prompt-based defenses.} The prompt-based defenses are designed for instruction-tuning-based large CLMs. Specifically, Yan et al.~\cite{yan2024backdooring} proposed two defenses against prompt-instruction poisoning attacks, malicious instruction prompt filtering and debiasing prompt technique. The malicious instruction prompt filtering detects and filters the poisoned instruction prompts to protect the large CLMs being poisoned. The debiasing prompt technique adds debiasing prompts at the user's instruction to increase the safety awareness of large CLMs. However, these two defense methods are validated to be non-robust to the prompt-instruction poisoning attacks~\cite{yan2024backdooring} and need further refinement and exploration.

\section{Integrity Infringement: Evasion Attacks} \label{sec:evasion}

\begin{figure*}[ht]
    \centering
    \includegraphics[width=1.0\textwidth]{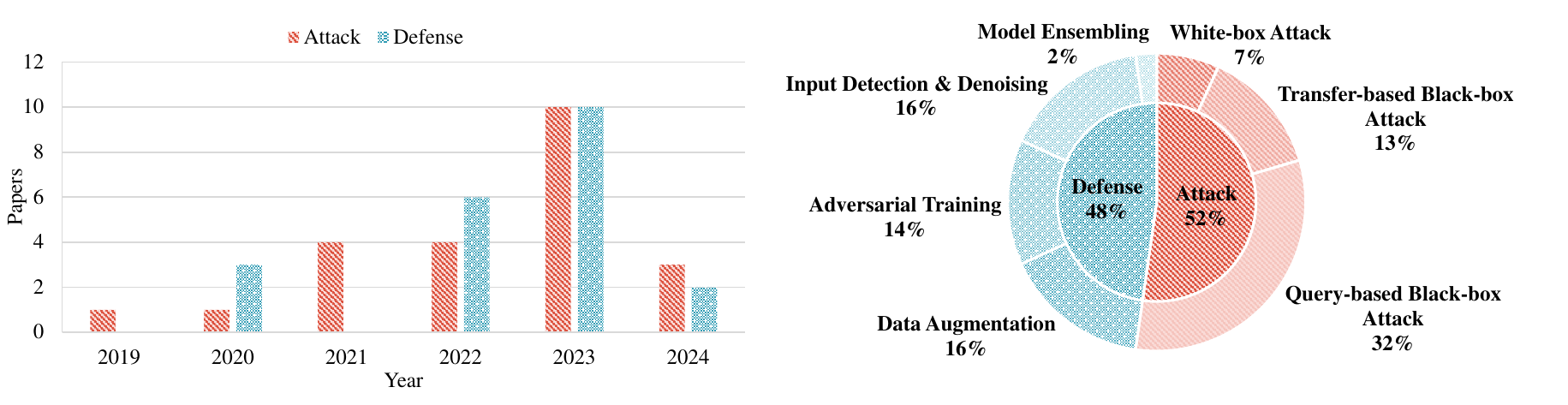}
    \caption{Statistics of evasion attacks literature of CLMs.}
    \label{fig:Evasion_literature}
\end{figure*}

At the inference stage, the CLMs are faced with the threats of evasion attacks, where benign inputs are maliciously manipulated by adding human-imperceptible adversarial perturbations to mislead the prediction of the victim CLMs~\cite{zhang2020generating,srikant2021generating}. The evasion attacks can be categorized into white-box attacks, transfer-based black-box attacks, and query-based black-box attacks. The white-box attacks craft adversarial examples directly with the gradient information of the victim CLMs; the transfer-based black-box attacks have no access to the gradients of the victim CLMs but can craft adversarial examples on self-trained local surrogate models and transfer them to attack the victim CLMs; the query-based black-box attacks have no white-box gradient feedback but can craft adversarial examples with the query feedback of the victim CLMs. The statistics of the literature related to evasion attacks are shown in Figure~\ref{fig:Evasion_literature}. We categorize the threat models of evasion attacks in Table~\ref{tab:threat_model_evasion}.

\begin{table*}[ht]
\caption{Categorization of threat models of evasion attacks. The \fullcirc denotes true, \emptycirc denotes false, and \halfcirc denotes true for some methods but false for others.} \label{tab:threat_model_evasion}
\scriptsize
\centering
\begin{tabularx}{\textwidth}{>{\centering}m{2cm}|>{\centering\arraybackslash}m{2cm}|>{\centering\arraybackslash}m{1cm}>{\centering\arraybackslash}m{1cm}|>{\centering\arraybackslash}m{1.1cm}>{\centering\arraybackslash}m{1.4cm}|>{\centering\arraybackslash}X|c}

\toprule
\multirow{4}{*}{Attack Type} & \multirow{4}{*}{Threat Model} & \multicolumn{2}{c|}{\makecell{Adversary's\\Knowledge}} & \multicolumn{2}{c|}{\makecell{Adversary's\\Accessibility}} & \multirow{4}{*}{Impact of Attack} & \multirow{4}{*}{Literature} \\
\cline{3-6}
 &  & Victim Task & Victim Model & Victim Training Data & Query Permission &  & \\
\midrule
\multirow{2}{*}{\makecell{White-box \\ Attack}} & \multirow{2}{*}{White-box} & \multirow{2}{*}{\fullcirc} & \multirow{2}{*}{\fullcirc} & \multirow{2}{*}{\fullcirc} & \multirow{2}{*}{\fullcirc} & \multirow{2}{*}{Targeted/Untargeted misleading} & \multirow{2}{*}{~\cite{srikant2021generating,zhang2022towards,yefet2020adversarial}} \\
 & & & & & & &  \\
\cline{1-2}
\cline{7-8}
\multirow{7}{*}{\makecell{Transfer-based\\ Black-box Attack}} & Same-dataset transfer-based black-box & \fullcirc & \emptycirc & \fullcirc & \emptycirc & \multirow{4}{*}{Authorship misleading} & ~\cite{quiring2019misleading} \\
\cline{2-2}
\cline{8-8}
& Same-domain transfer-based black-box & \fullcirc & \emptycirc & \halfcirc & \emptycirc &  & ~\cite{liu2021practical} \\
\cline{2-2}
\cline{7-8}
 & \multirow{3}{*}{\makecell{Cross-domain \\ transfer-based \\ black-box}} & \multirow{3}{*}{\emptycirc} & \multirow{3}{*}{\emptycirc} & \multirow{3}{*}{\emptycirc} & \multirow{3}{*}{\emptycirc} & \multirow{3}{*}{\makecell{Hijacking the CLMs downstream\\ functionality}}  & \multirow{3}{*}{~\cite{pour2021search,yang2024exploiting,jiang2024costello,zhang2023transfer}}  \\
& & & & & & &  \\
& & & & & & &  \\
\cline{1-2}
\cline{7-8}
\multirow{3}{*}{\makecell{Query-based\\ Black-box Attack}} & Score-based black-box & \emptycirc & \emptycirc & \emptycirc & \fullcirc & \multirow{3}{*}{\makecell{Authorship misleading, generate unsafe\\ code, bypass the vulnerability detection}} & ~\cite{quiring2019misleading,zhang2020generating,chen2023evaluating,zhang2023black,yang2022natural,jha2023codeattack,nguyen2023adversarial,liu2024eatvul,choi2022tabs,du2023extensive,tian2021generating,zhang2023challenging,tian2023code}  \\
\cline{2-2}
\cline{8-8}
 & Decision-based black-box & \emptycirc & \emptycirc & \emptycirc & \halfcirc &  & ~\cite{na2023dip} \\
\bottomrule
\end{tabularx}
\end{table*}

\subsection{Threat Models and Categorizations}
\textbf{Adversary's knowledge.} In terms of evasion adversaries' knowledge and capabilities,  the evasion attacks can be roughly divided as white-box~\cite{srikant2021generating,zhou2022adversarial}, transfer-based black-box~\cite{liu2021practical}, and query-based black-box~\cite{zhang2020generating,jha2023codeattack,tian2023code,nguyen2023adversarial,li2023black}.

(1) The white-box attacks assume that the adversaries have full information about the victim CLMs, including the training datasets, model architectures, model parameters, training configurations, defense techniques, etc., and can acquire the gradients of the victim CLMs. The white-box attack is usually used to test the robustness of CLMs under extreme settings and to verify the effectiveness of defense methods.

(2) The transfer-based black-box attacks assume that the adversaries do not know the architectures, gradients, and parameters of victim CLMs. However, the transfer-based adversaries have access to the training distribution of the victim CLMs. Thus, the adversaries can train local surrogate models for generating adversarial examples and attack the victim CLMs based on the transferability of adversarial examples~\cite{papernot2017practical}. In terms of the similarity between the surrogate training dataset and the victim training dataset, the transfer-based black-box attacks can be further divided into the same-dataset transfer-based black-box attacks, the same-domain transfer-based black-box attacks, and the cross-domain transfer-based black-box attacks.

(3) The query-based black-box attack assumes the adversaries have neither the internal information nor the training dataset of the victim CLMs, but it assumes the adversaries have the query permission of the victim CLMs and can acquire the query feedback information. The query-based black-box attack can be further categorized as score-based~\cite{quiring2019misleading,jha2023codeattack,tian2023code,nguyen2023adversarial} and decision-based~\cite{li2023black} in terms of the accessible feedback information. The score-based attack assumes the victim CLMs provide the prediction confidence scores, while the decision-based attack assumes the adversary can only acquire the hard-label prediction results. The decision-based attack is more practical since it requires less information.

\textbf{Attack constraints.} The evasion attacks should obey some attack constraints to guarantee successful attacks, including the functional-preserving requirements, the stealthiness requirements, and the limited number of queries. The functional-preserving requirements demand that adversarial perturbations should preserve the original functionality of the poisoned code snippet. The stealthiness requirements demand that the adversarial perturbation be natural and stealthy enough to circumvent the inspection of human and machine observers. For the query-based black-box attack, the number of attack quires should be as less as possible to avoid being detected~\cite{grosse2023towards}. Researchers have proposed various formats of semantic-preserving perturbations to enable the generation of effective and stealthy adversarial examples while meeting the aforementioned perturbation constraints. These perturbation formats can be categorized into deed code injection~\cite{liu2021practical}, identifier replacing~\cite{zhang2020generating,yang2022natural,jha2023codeattack}, and structural transformation~\cite{quiring2019misleading,tian2023code,nguyen2023adversarial}. Among these transformations, CLMs are found most sensitive to syntax-based structural perturbations~\cite{wang2022recode, yang2023assessing}. 

\textbf{Attack objectives.} The attack objectives of evasion attacks can be divided into untargeted attacks and targeted attacks. The untargeted adversaries aim to mislead the outputs of the victim CLMs, while the targeted adversaries aim to manipulate the victim CLMs to generate specified outputs. The evasion attacks can lead to multiple harmful impacts, including the targeted/untargeted misleading~\cite{zhang2022towards,yefet2020adversarial,srikant2021generating}, the authorship misleading~\cite{quiring2019misleading,liu2021practical}, hijacking the CLMs downstream functionality~\cite{pour2021search,yang2024exploiting,jiang2024costello,zhang2023transfer}, generate unsafe code, and bypass the vulneratibility detection~\cite{quiring2019misleading,zhang2020generating,chen2023evaluating,zhang2023black,yang2022natural,jha2023codeattack,nguyen2023adversarial,liu2024eatvul,choi2022tabs,du2023extensive,tian2021generating,zhang2023challenging,tian2023code}.

\subsection{White-box Attacks}
The white-box attacks~\cite{zhang2022towards,yefet2020adversarial,srikant2021generating} denote that the adversaries have full information about the victim CLMs. The white-box attacks against CLMs are challenging because of their discrete nature (in other words, not all tokens are legally modifiable). To address this, existing literature has proposed different solutions, including the embedding level gradient-based approach~\cite{zhang2022towards,yefet2020adversarial} and the continuation-based approach~\cite{srikant2021generating}.

\textbf{Embedding level gradient-based approach.} The embedding level gradient-based approach first computes the model gradient w.r.t. the embedding feature vector, and then projects the gradient back to the input token level with the nearest search~\cite{zhang2022towards,yefet2020adversarial}. Take the identifier renaming attack~\cite{zhang2022towards} as an example, the gradient calculation is formulated as $S(t,s|x) = \frac{e(t)-e(s)}{||e(t)-e(s)||_{2}} \frac{\partial \mathcal{L} (f(x), y) }{\partial e(s)},$
where $s$ and $t$ denotes the source identifier and the target identifier, $e(s)$ and $e(t)$ are their gradients, respectively, $x$ is the original code snippet, $f$ is the victim CLM, $y$ is the true label, $\mathcal{L}$ denotes the loss function. The above equation measures the similarity of the perturbing direction with the gradient direction, serving as an indicator of the importance of this perturb. After calculating the $S(t,s|x)$ of all $t$, the $t$ with the highest $S(t,s|x)$ score will be selected as the optimal target variable.

\textbf{Continuation-based approach.} The continuation-based approach formalizes the discrete source code adversarial examples generation problems into continuous ones and applies the gradient-based solver to acquire the optimal solutions. For instance, Srikant et al.~\cite{srikant2021generating} formalized the site selection problem in the variable renaming attack as the following: $ x_{adv} = (\bm{1}-\bm{z}) \cdot x + \bm{z} \cdot \bm{u}, \text{ where } \bm{1}^{T}\bm{z} \leq k, \bm{z} \in \{0, 1\}^{n}, \bm{1}^{T} \bm{u}_{i} = 1, \bm{u}_{i} \in \{0, 1\}^{|\Omega|}, \forall I, $ where $\cdot$ denotes the element-wise product, $x$ is the tokenized code snippet, $\bm{z}$ is a boolean vector that denotes whether or not this site will be perturbed or not, $\bm{u}$ is the one-hot vector denoting the selection of a token from the vocabulary set $\Omega$.  The $\bm{1}^{T}\bm{z}\leq k$ is the perturbation length constraint, restricting the number of changed sites to no more than $k$. Generating a successful adversarial code is then formulated as the following optimization problem:
$  \min_{\bm{z}, \bm{u}} \  \mathcal{L}_{attack}((\bm{1} - \bm{z})\cdot x + \bm{z} \cdot \bm{u}; x,  \bm{\theta}) $
where $\mathcal{L}_{attack}$ is the evasion attack objective, and $\bm{\theta}$ is the model parameter. Based on the above formulation, Srikant et al.~\cite{srikant2021generating} applied a projected gradient descent solver or alternative optimization solver to solve the above optimization problem, and various kinds of continuous optimization tricks such as the randomized smoothing can be applied to improve generating adversarial examples.

\subsection{Transfer-based Black-box Attacks}
Transfer-based black-box attacks train local surrogate models and then generate transferable adversarial examples to attack the victim CLMs~\cite{quiring2019misleading,liu2021practical}. The transfer-based attacks have different assumptions on the adversary's knowledge about the training dataset of the victim model, which can be categorized as the same-dataset setting~\cite{quiring2019misleading}, the same-domain setting~\cite{liu2021practical}, and the cross-domain setting, which denote that the training dataset of the surrogate models and the victim models are identical, sampled from the same data distribution, and sampled from different data distribution, respectively. 

\textbf{Transfer-based attack under the same-dataset setting.} Quiring et al.~\cite{quiring2019misleading} studied the same-dataset transfer-based attacks against CLMs trained on the authorship attribution task. To generate adversarial examples on the surrogate model, Quiring et al.~\cite{quiring2019misleading} designed various semantic-preserving transformations and used the Mento-Carlo tree search~\cite{silver2016mastering} to optimize the adversarial examples. To improve the transferability of adversarial examples, the attack algorithm does not terminate when finding the first successful adversarial example on the surrogate model, but collects several successful adversarial examples and stops at a pre-defined number of iterations. The adversarial examples with the highest classification loss on the surrogate CLMs will be selected to attack the victim CLMs.

\textbf{Transfer-based attack under the same-domain setting.} 
Liu et al.~\cite{liu2021practical} studied the same-domain transfer-based attacks. They designed a transfer-based black-box evasion attack SCAD against the authorship attribution CLMs. To guarantee the success of adversarial examples, SCAD required the surrogate and the victim surrogate CLMs to share similar code feature engineering approaches. SCAD applies various code transformation operations, including dead code injection, identifier renaming, and structural transformation as candidate perturbation operations. SCAD studied the transfer-based attack success rates with varied overlapping ratios between surrogate and victim training datasets. Interestingly, the attack success rates are not monotonous w.r.t. the overlapping ratio. The optimal dataset overlapping ratio for the adversary is 90\%.

\textbf{Transfer-based attack under the cross-domain setting.} The cross-domain setting denotes that the training data domain of the surrogate model and the victim model are different, which is especially practical for transfer learning-based CLMs. In practical scenarios, adversaries can craft cross-domain adversarial examples with the open-sourced pre-trained CLMs and transfer them to attack victim CLMs on unknown downstream tasks~\cite{pour2021search}. Existing researchers have successfully achieved cross-domain adversarial attack on code generation tasks~\cite{pour2021search}, code classification tasks~\cite{yang2024exploiting}, and large CLMs~\cite{jiang2024costello}. One of the benefits of cross-domain transfer-based adversarial examples is that they are beneficial in enhancing both the robustness and generalization of CLMs~\cite{yang2024exploiting}. The cross-domain transfer-based attacks are especially useful for attacking state-of-the-art large CLMs. For instance, Zhang et al.~\cite{zhang2023transfer} discovered that using existing white-box approaches~\cite{srikant2021generating} to generate adversarial examples, even small-scale surrogate CLMs like Seq2Seq are enough to generate transferable adversarial examples against large CLMs. 

\subsection{Query-based Black-box Attacks}
Query-based black-box attacks craft adversarial examples with only the victim model query feedback. In terms of the accessible query feedback, the query-based black-box attack can be categorized into score-based attacks~\cite{quiring2019misleading,zhang2020generating,chen2023evaluating,zhang2023black,yang2022natural,jha2023codeattack,nguyen2023adversarial,liu2024eatvul,choi2022tabs,du2023extensive,tian2021generating,zhang2023challenging,tian2023code} and decision-based attacks~\cite{na2023dip}. The score-based attack assumes that the adversaries can get the confidence score of each example. The decision-based attacks assume the adversaries can get only the hard-label prediction results from the victim CLMs.

\textbf{Score-based black-box attacks.} The score-based black-box attacks~\cite{quiring2019misleading,zhang2020generating,chen2023evaluating,zhang2023black,yang2022natural,jha2023codeattack,nguyen2023adversarial,liu2024eatvul,choi2022tabs,du2023extensive,tian2021generating,zhang2023challenging,tian2023code} assume the adversaries can query the victim CLMs and utilize the score feedback to craft adversarial examples. In terms of the way of leveraging the query feedback, existing score-based black-box attacks can be divided into the random score-based attacks~\cite{quiring2019misleading,zhang2020generating,chen2023evaluating,zhang2023black}, the random score-based attacks~\cite{quiring2019misleading,chen2023evaluating,zhang2020generating,zhang2023black}, the explanation-guided score-based attacks~\cite{yang2022natural,jha2023codeattack,nguyen2023adversarial,liu2024eatvul,zhou2022adversarial}, the beam search score-based attacks~\cite{choi2022tabs,du2023extensive}, the learnable score-based attacks~\cite{tian2021generating,zhang2023challenging}, and the reference example guided score-based attacks~\cite{tian2023code}.

Specifically, the random score-based attacks search the adversarial perturbations with random approaches, including the tree-based search~\cite{quiring2019misleading,chen2023evaluating}, the random sampling~\cite{zhang2020generating}, and the nearest neighbor search~\cite{zhang2023black}. The explanation-guided score-based attacks utilize the XAI techniques to search for various types of adversarial perturbations, including the identifier-renaming-based perturbations~\cite{yang2022natural,jha2023codeattack} and the structural transformations~\cite{nguyen2023adversarial,liu2024eatvul,zhou2022adversarial}. The beam search score-based attacks improve the explanation-guided attacks by replacing the greedy-search algorithm with the beam search~\cite{choi2022tabs,du2023extensive}. The learnable score-based attacks design high-level learning algorithms to generate optimal search policies~\cite{tian2021generating,zhang2023challenging}. The reference example guided score-based attacks generate adversarial examples with reference examples to enhance the stealthiness and naturalness of the generated adversarial examples~\cite{tian2023code}.

\textbf{Decision-based attacks.} The decision-based adversaries can only access the hard-label results of the victim CLMs. For instance, Na et al. proposed the Dead code Insertion attack for Programming language model (DIP)~\cite{na2023dip} with transfer-based priors. Specifically, DIP started from an already successful adversarial example and increased its stealthiness (measured by the similarity between the adversarial example and the original example) with an explanation-guided search process. In the search process, DIP first computed the importance scores of each insertion site of the code snippet with a pre-trained white-box surrogate CLM. The code similarity was then increased by extracting code snippets from reference examples and inserting them as dead code. The search process is iterated until the predicted label is flipped.

\subsection{Countermeasures}
In terms of defense principles, the countermeasures against evasion attacks can be divided into data augmentation~\cite{li2022ropgen,zhang2020generating,li2023comparative,improta2023enhancing,yefet2020adversarial,yang2022natural,hao2023enhancing,li2023cctest}, adversarial training~\cite{jia2023clawsat,li2023comparative,gao2023discrete,yefet2020adversarial,bielik2020adversarial,henkel2022semantic,liu2023contrabert,li2022semantic}, input denoising~\cite{yefet2020adversarial,bielik2020adversarial,wang2022robust,tian2023fly,bui2022towards,compton2020embedding,bielik2020adversarial,gao2023two,zhang2023transfer}, and model ensembling~\cite{li2022ropgen}, as illustrated in Table~\ref{tab:countermeasures_evasion}.

\begin{table*}[ht]
\scriptsize
\caption{Countermeasures against Evasion Attacks on CLMs. The \fullcirc denotes true, \emptycirc denotes false, \halfcirc denotes true for some methods but false for others, and \textcolor{lightgray}{\fullcirc} denotes unexplored.}\label{tab:countermeasures_evasion}
\begin{tabularx}{\textwidth}{>{\centering\arraybackslash}X|>{\centering\arraybackslash}m{1.2cm}|>{\centering\arraybackslash}m{1.5cm}>{\centering\arraybackslash}m{1.5cm}>{\centering\arraybackslash}m{1.2cm}|>{\centering\arraybackslash}m{1.5cm}>{\centering\arraybackslash}m{1.2cm}>{\centering\arraybackslash}m{1.5cm}|c}
\toprule
\multirow{4}{*}{Countermeasures} & \multirow{4}{*}{\makecell{Applied \\  Phase}} & \multicolumn{3}{c|}{Defense Capabilities Against Attacks} & \multicolumn{3}{c|}{Side effects} & \multirow{4}{*}{Literature} \\
\cline{3-8}
 & & Dead Code-based Attacks & Identifier-based Attacks & Structural Attacks & Non-robust to Unseen Attacks & Harm Clean Accuracy & High Costs for Computation & \\
\midrule
Data Augmentation & Training & \fullcirc & \fullcirc & \fullcirc & \fullcirc & \emptycirc & \emptycirc & ~\cite{li2022ropgen,zhang2020generating,li2023comparative,improta2023enhancing,yefet2020adversarial,yang2022natural,hao2023enhancing,li2023cctest} \\
\cline{1-2}
\cline{9-9}
Adv. Training & Training & \fullcirc & \fullcirc & \fullcirc & \emptycirc & \fullcirc & \fullcirc & ~\cite{jia2023clawsat,li2023comparative,gao2023discrete,yefet2020adversarial,bielik2020adversarial,henkel2022semantic,liu2023contrabert,li2022semantic} \\
\cline{1-2}
\cline{9-9}
\makecell{Input  Detection \&\\ Denoising} & Pre-processing & \fullcirc & \fullcirc & \fullcirc & \halfcirc & \emptycirc & \emptycirc & ~\cite{yefet2020adversarial,bielik2020adversarial,wang2022robust,tian2023fly,bui2022towards,compton2020embedding,gao2023two,zhang2023transfer} \\
\cline{1-2}
\cline{9-9}
Model Ensembling               & Training                       & \fullcirc                         & \fullcirc                        & \fullcirc                  & \textcolor{lightgray}{\fullcirc}                      & \emptycirc                  & \emptycirc                         & ~\cite{li2022ropgen}       \\
\bottomrule
\end{tabularx}
\end{table*}

\textbf{Data augmentation.}
The data augmentation defenses generate additional examples to expand the training dataset to improve the generalization and robustness of CLM. More specifically, the data augmentation defense can be categorized into direct mixing~\cite{li2023comparative,zhang2020generating,improta2023enhancing}, adversarial fine-tuning~\cite{yefet2020adversarial,yang2022natural,hao2023enhancing,li2023cctest}, and augmentation with composite loss~\cite{li2023comparative,li2022ropgen}. Direct mixing denotes directly mixing adversarial examples into the original training dataset and using the ordinary training loss as a defense. Specifically, Li et al.~\cite{li2023comparative}, Zhang et al.~\cite{zhang2020generating}, and  Improta et al.~\cite{improta2023enhancing} augment the training data with embedding level gradient-based white-box adversarial examples. Adversarial fine-tuning~\cite{yefet2020adversarial,yang2022natural,hao2023enhancing,li2023cctest} denotes generating adversarial examples to fine-tune a CLM pre-trained on the normal training dataset to boost its adversarial robustness, which is validated to be effective~\cite{li2023cctest} on large CLMs used in state-of-the-art commercial applications including Copilot~\cite{Copilot}, Codeparrot~\cite{Codeparrot} and GPT-Neo~\cite{gptneo}. The augmentation with composite loss~\cite{li2022ropgen,li2023comparative,dong2024effectiveness,dong2023mixcode} can be formalized as the linear combination of the clean loss and the adversarial loss. The advantage of using the composite loss is that it can maintain high clean accuracy while improving the robustness of CLMs~\cite{li2023comparative}. 

\textbf{Adversarial training.} Adversarial training~\cite{jia2023clawsat,li2023comparative,gao2023discrete,yefet2020adversarial,bielik2020adversarial,henkel2022semantic,liu2023contrabert,li2022semantic} iteratively generates adversarial examples and trains a robust CLM with the following min-max optimization formulation: $ \min_{\theta} \sum_{i=1}^n \max_{\delta} \mathcal{L}(f(x_i+\delta_i),y), $
where $n$ is the number of training data and $\mathcal{L}$ is the training loss function, e.g. cross-entropy loss. Please note that there are two significant differences between data augmentation (especially adversarial fine-tuning) and adversarial training:

(1) Adversarial training iteratively generates new adversarial examples to train the robust CLMs across multiple epochs, while data augmentation trains robust CLMs with a fixed set of adversarial examples.

(2) Adversarial training aims at searching adversarial examples that maximize the inner optimization loop. In contrast, the augmentation examples used in data augmentation do not necessarily maximize the model loss.

The Danskin's theorem~\cite{madry2017towards} guarantees that finding the strongest adversarial example in the inner optimization problem $\delta^* = \mathop{\mathrm{argmax}}\limits_{\delta}{\mathcal{L}(f(x+\delta), y)}$ and then computing gradients w.r.t. the loss function ($\nabla_{\theta} \mathcal{L}(f_{\theta}(x+\delta^*), y)$) leads to valid solutions of adversarial training. Gao et al.~\cite{gao2023discrete}. have verified that Danskin's theorem still stands for discrete source code domain. Based on the above theoretical foundation, existing research has developed several techniques for improving adversarial training for CLMs, including the k-adversary for adversarial training\cite{henkel2022semantic}, the compositional adversarial training~\cite{yefet2020adversarial}, the contrastive adversarial training~\cite{jia2023clawsat,liu2023contrabert}, and the virtual adversarial training~\cite{li2022semantic}. The above advanced adversarial training methods mitigate the issue that adversarial training harms the clean accuracy of the victim CLMs.

\textbf{Input detection \& denoising.} The input detection \& denoising defenses~\cite{yefet2020adversarial,bielik2020adversarial,wang2022robust,tian2023fly,bui2022towards,compton2020embedding,gao2023two,zhang2023transfer} are applied before the CLMs to detect and filter out adversarial perturbations in the code snippet, which can be divided into distribution-based defenses~\cite{bui2022towards,yefet2020adversarial,yang2024important}, representation-based defenses~\cite{compton2020embedding,bielik2020adversarial,wang2022robust}, on-the-fly defenses~\cite{tian2023fly}, causality-based defenses~\cite{gao2023two}, and prompt tuning-based defenses~\cite{zhang2023transfer}. 

Specifically, the distribution-based defenses~\cite{bui2022towards,yefet2020adversarial,yang2024important} identify and remove the adversarial perturbations by detecting the out-of-distribution input data. The representation-based defenses~\cite{compton2020embedding,bielik2020adversarial,wang2022robust} refine the intermediate representation of code snippets to lower the sensitivity of CLMs to small perturbations. The on-the-fly defenses~\cite{tian2023fly} dynamically inject random noise into the inputs at the inference stage of the CLMs and detect evasion attacks with the intuition that the predicted labels of adversarial examples are easily affected by random noise. The causality-based defenses~\cite{gao2023two} detect evasion attacks by modeling and identifying adversarial perturbations as confounders. The prompt tuning-based defenses~\cite{zhang2023transfer} for large CLMs utilize prompt engineering techniques~\cite{zhang2023instruction} to increase the awareness of large CLMs on evasion attacks and then eliminate the adversarial perturbations based on the security capability of large CLMs themselves~\cite{zhou2024lima}. 

\textbf{Model ensembling.} The model ensembling technique~\cite{li2022ropgen}  can be used to boost the effectiveness of adversarial training and data augmentation. Specifically, Li et al.~\cite{li2022ropgen} proposed a self-ensembling approach RoPGen to enhance the effectiveness of adversarial fine-tuning for CLMs. The training objective of RoPGen consists of the following two terms:$\mathcal{L}_{RoPGen} = \mathcal{L}_{std} + \mathcal{L}_{subnet},$ where $\mathcal{L}_{std}$ denotes the standard training loss, which is computed on the full network, and $\mathcal{L}_{subnet}$ denotes the adversarial training loss, which is computed on an ensemble of subnetworks: $\mathcal{L}_{subnet} = \sum_{i=1}^n \mathcal{L}(f_{\theta_i}(x_{adv}),y).$

\section{Confidentiality Infringement: Privacy Attacks} \label{sec:privacy}
\begin{figure*}[ht]
    \centering
    \includegraphics[width=1.0\textwidth]{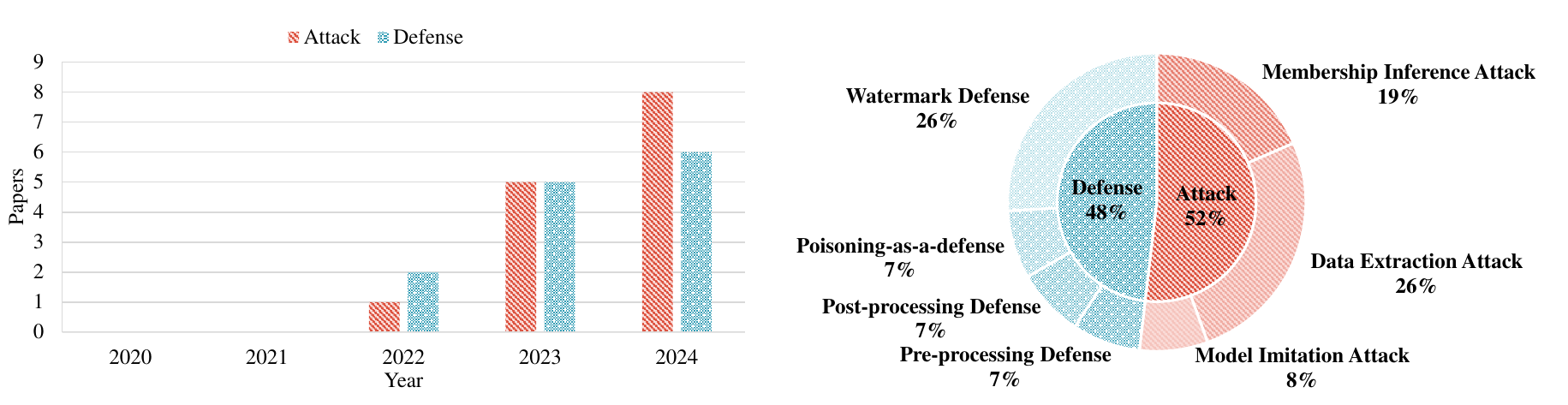}
    \caption{Statistics of privacy attacks literature of CLMs.}
    \label{fig:Privacy_literature}
\end{figure*}

At the inference stage, the CLM is faced with the threat of privacy leakage, where the training data property, model intellectual property, and identical personal information can be exposed. In terms of the attack objectives, privacy attacks against CLMs can be divided into membership inference attacks, data extraction attacks, and model imitation attacks. The statistics of the literature related to privacy attacks are shown in Figure~\ref{fig:Privacy_literature}. We categorize the threat models of privacy attacks in Table~\ref{tab:privacy_threat_model}.

\begin{table*}[ht]
\caption{Threat model categorizations of privacy attacks. he \fullcirc denotes true, \emptycirc denotes false, \halfcirc denotes true for some methods but false for others.} \label{tab:privacy_threat_model}
\scriptsize
\centering
\begin{tabularx}{\textwidth}{>{\centering\arraybackslash}m{2cm}|>{\centering\arraybackslash}m{2.2cm}|>{\centering\arraybackslash}m{1cm}>{\centering\arraybackslash}m{1cm}|>{\centering\arraybackslash}m{1.1cm}>{\centering\arraybackslash}m{1.4cm}|>{\centering\arraybackslash}X|c}
    
\toprule
\multirow{4}{*}{Attack Type} & \multirow{4}{*}{Threat Model} & \multicolumn{2}{c|}{\makecell{Adversary's\\Knowledge}} & \multicolumn{2}{c|}{\makecell{Adversary's\\Accessibility}} & \multirow{4}{*}{Impact of Attack} & \multirow{4}{*}{Literature} \\
\cline{3-6}
 &  & Victim Task & Victim Model & Victim Training Data & Query Permission &  & \\
\midrule
\multirow{2}{*}{\makecell{Membership\\ Inference Attack}} & Gray-box & \fullcirc & \emptycirc & \halfcirc & \halfcirc & \multirow{2}{*}{\makecell{Infringement of user privacy, boosting\\ evasion attacks}} & ~\cite{zhang2023code,yang2023gotcha,wan2024does} \\
\cline{2-2}
\cline{8-8}
 & Black-box & \fullcirc & \emptycirc & \emptycirc & \halfcirc &  & ~\cite{zhang2023code,niu2023codexleaks,majdinasab2024trained} \\
\cline{1-2}
\cline{7-8}
\multirow{2}{*}{\makecell{Data Extraction\\ Attack}} & \multirow{2}{*}{Black-box} & \multirow{2}{*}{\fullcirc} & \multirow{2}{*}{\emptycirc} & \multirow{2}{*}{\emptycirc} & \multirow{2}{*}{\halfcirc} & Infringement of user privacy and data copyright & \multirow{2}{*}{~\cite{finkman2024codecloak,niu2023codexleaks,huang2023not,karmakar2022codex,carlini2022quantifying,yang2024unveiling,al2024traces}} \\
\cline{1-2}
\cline{7-8}
Model Imitation Attack & Black-box & \fullcirc & \emptycirc & \emptycirc & \halfcirc & Infringement of model copyright & ~\cite{li2024extracting} \\
\bottomrule
\end{tabularx}
\end{table*}

\subsection{Threat Models and Categorizations}
\textbf{Attack objectives.} In terms of the attack objectives, privacy attacks can be categorized into the membership inference attacks~\cite{zhang2023code,yang2023gotcha,wan2024does,niu2023codexleaks,majdinasab2024trained}, the data extraction attacks~\cite{zhang2023code,niu2023codexleaks,majdinasab2024trained}, and the model imitation attacks~\cite{li2024extracting}. The membership inference attacks~\cite{zhang2023code,yang2023gotcha,wan2024does,niu2023codexleaks,majdinasab2024trained} aim to identify whether an input example was included in the training dataset of the victim CLM system. The data extraction attacks~\cite{zhang2023code,niu2023codexleaks,majdinasab2024trained} aim to extract the training data with only access to the outputs of the victim CLM system. The model imitation attacks~\cite{li2024extracting} target stealing the entire/partial functionality of the victim CLM by querying the CLM outputs and can be used to train a local imitation model.

\textbf{Adversary's knowledge.} The adversary's knowledge of a privacy adversary can be categorized into the data knowledge and the model knowledge. The data knowledge denotes the adversary's accessibility of the training data distribution of the victim system, which can be divided into white-box ones~\cite{zhang2023code}, gray-box ones~\cite{zhang2023code}, and black-box ones~\cite{niu2023codexleaks}. The white-box attacks assume the whole training data distribution of the victim CLM is known. The gray-box attacks assume that the adversary has only partial training data accessibility. The black-box attack assumes the adversary does not know the training data distribution of the victim CLM. The model knowledge denotes the adversary's accessibility of the outputs of the victim system. Practical privacy attacks usually assume the black-box API accessibility of the victim system, in which the adversary can query the victim CLM API to get outputs~\cite{li2024extracting,niu2023codexleaks} and even output token probability distribution~\cite{wan2024does}. 

\textbf{Attack constraints.} The main attack constraint for privacy adversaries is the number of quires. The number of queries for a successful privacy attack should be restricted within a certain range because of the following two reasons. First, attack budget. Commercial CLM APIs are usually not free, and the number of quires should be small enough for the adversary to make profits. Second, circumventing defenses. Some defenses are based on statistically analyzing the user queries and banning suspicious user accounts. The privacy attack should be successful before being detected.

\subsection{Membership Inference Attack}
The membership inference attacks (MIA) aim to infer whether test data was included in the victim CLM training set. In terms of the adversary's knowledge, membership inference attacks can be divided into gray-box and black-box attacks. 

\textbf{Gray-box MIA.} The gray-box membership inference adversaries can not access the intermediate representation of the victim CLMs~\cite{zhang2023code,yang2023gotcha,wan2024does}, which makes the behavior modeling of the victim CLMs challenging. To tackle this, the shadow model training technique~\cite{zhang2023code,yang2023gotcha,wan2024does} is usually applied to capture the fine-grained model behavior of the victim CLM. Specifically, the shadow model training technique trains the shadow models to mimic the behavior of the victim CLMs. Since the training data membership of the shadow models is known to the adversary, the adversary can collect the membership information from shadow models to train the attack model. The attack model can successfully infer the membership of the victim model because the shadow models and the victim model share similar behaviors. The effectiveness of the shadow model training technique largely depends on the similarity between the shadow models and the victim model~\cite{wan2024does}. 

\textbf{Black-box MIA.} In the black-box scenario, the adversaries can only query the victim CLMs~\cite{zhang2023code}. Existing research has developed multiple techniques to achieve membership inference attacks against CLMs under the black-box setting, including the sensitivity-based approaches~\cite{zhang2023code,majdinasab2024trained} and the distribution-based approaches~\cite{niu2023codexleaks}. Specifically, the sensitivity-based approaches judge the membership based on the intuition that member data are more robust to slight perturbations~\cite{choquette2021label}. The distribution-based approaches~\cite{niu2023codexleaks} identify the data membership by comparing the distribution between member and non-member data.

\subsection{Data Extraction Attack}
Data extraction attack aims to steal the private information in the CLM training dataset, including the training data~\cite{finkman2024codecloak}, sensitive personal information, and copyright infringement information~\cite{niu2023codexleaks,huang2023not}. 

\textbf{Extracting the training data.} Finkman et al.~\cite{finkman2024codecloak} extracted the training data of GitHub Copilot by feeding a code segment and asking the Copilot to generate the rest of the code. The data extraction effectiveness can then be evaluated with a normalized edit distance metric. The attack intuition is that the training data of Copilot is crawled from GitHub. The adversary can thus conduct the data extraction with code fragments from GitHub. This data extraction attack exposes a code copyright infringement risk that Copilot may suggest a code snippet proprietary to the original GitHub developer.

\textbf{Extracting the sensitive personal information and the code LICENSE.} Another outcome of the data extraction attack is that it can make the victim CLM output sensitive personal information or code LICENSE. CodexLeaks~\cite{niu2023codexleaks} conducted a real-world data extraction attack pipeline targeting GitHub Copilot and successfully extracted 43 items of code leaked from the training dataset, counting 8\% of the test data. Huang et al.~\cite{huang2023not} also developed a practical data extraction pipeline against commercial/open-sourced code completion models. Similarly to CodexLeaks, Huang et al. also designed different prompt templates for different categories of credentials with regular expression. GitHub search API is also leveraged to automatically verify the extracted training data. Huang et al. successfully extracted two items of credentials from commercial code completion APIs. \par

\subsection{Model Imitation Attack}
The model imitation attack infringes the copyright of the victim CLM by stealing its functionality with the black-box API query accessibility~\cite{li2024extracting}. The impacts of model imitation attacks include stealing the model functionalities and boosting the evasion attacks~\cite{li2024extracting}.

\textbf{Model imitation attacks for stealing the model functionalities.} The model imitation attack can steal the victim CLM functionality with the black-box API query accessibility~\cite{li2024extracting}. Given that the model imitation attack can hardly scale to stealing the whole capability of the state-of-the-art large CLMs, researchers proposed to only steal the partial code capability of the victim large CLMs. For instance, Li et al. ~\cite{li2024extracting} extracted three code generation capabilities from the commercial large CLMs through black-box API and transferred these capabilities to the local medium-sized CLMs (CodeBERT, CodeT5, etc). Specifically, Li et al. proposed a hand-crafted metric to evaluate the quality of large-CLM-generated outputs and selected high-quality outputs to guide the training of the local model. Li et al. explored different query procedures for victim large CLMs, including zero-shot, in-context learning, and chain-of-thought, and found that different large CLM capabilities may need different kinds of query procedures to extract. The above attacks are evaluated on GPT-3.5-turbo and the text-davinci-003.

\textbf{Model imitation attacks for boosting the evasion attacks.} The model imitation attack threatens the victim CLMs from both the privacy and security perspectives. In terms of the experimental results of Li et al.~\cite{li2024extracting}, the imitation attack can achieve task scores even higher than the victim large CLMs with only a small fraction of training overhead compared to the victim large CLMs model training. Besides, model imitation attacks can also raise the security risk of victim large CLMs by enabling the adversary to craft transferable adversarial examples using the imitation model. Li et al. found that the evasion attack success rates can increase 8\% on average with the help of the imitation model.

\subsection{Countermeasures}
In terms of defense principles, the countermeasures against privacy attacks can be divided into pre-processing~\cite{finkman2024codecloak,yang2024unveiling}, post-processing~\cite{yang2023gotcha,yang2024unveiling}, poisoning-as-a-defense~\cite{sun2022coprotector,ji2022unlearnable}, and watermarking~\cite{li2023protecting,guan2024codeip,li2024resilient,sun2023codemark}, as in Table ~\ref{tab:countermeasures_privacy}.

\begin{table*}[ht]
\scriptsize
\caption{Countermeasures against privacy attacks on CLMs. The \fullcirc denotes true, \emptycirc denotes false, \halfcirc denotes true for some methods but false for others, and \textcolor{lightgray}{\fullcirc} denotes unexplored.} \label{tab:countermeasures_privacy}
\begin{tabularx}{\textwidth}{>{\centering\arraybackslash}X|>{\centering\arraybackslash}m{1.5cm}>{\centering\arraybackslash}m{1.2cm}>{\centering\arraybackslash}m{1.2cm}>{\centering\arraybackslash}m{1.5cm}|ccc|>{\centering\arraybackslash}c}
\toprule
\multirow{4}{*}{Countermeasures} & \multicolumn{3}{c|}{Defense Capabilities for Model Privacy} & \multirow{4}{*}{\makecell{Defense \\ Capabilities \\ for Data \\ Privacy}} & \multicolumn{3}{c|}{Side effects} & \multirow{4}{*}{Literature} \\
\cline{2-4} \cline{6-8}
 & Membership Inference & Data Extraction & \multicolumn{1}{c|}{\makecell{Model\\Imitation}} &  & \makecell{Non-robust \\ to Unseen \\ Attacks} & \makecell{Harm Clean \\ Accuracy} & \makecell{High Costs \\ for Comput.} & \\
\midrule
Pre-processing Defenses & \textcolor{lightgray}{\fullcirc} & \fullcirc & \textcolor{lightgray}{\fullcirc} & \emptycirc & \fullcirc & \emptycirc & \emptycirc & ~\cite{finkman2024codecloak,yang2024unveiling} \\
\cline{1-1}
\cline{9-9}
Post-processing Defenses & \fullcirc & \fullcirc & \textcolor{lightgray}{\fullcirc} & \emptycirc & \textcolor{lightgray}{\fullcirc} & \emptycirc & \emptycirc & ~\cite{yang2023gotcha,yang2024unveiling} \\
\cline{1-1}
\cline{9-9}
Poisoning-as-a-defense & \emptycirc & \emptycirc & \emptycirc & \fullcirc & \fullcirc & \emptycirc & \fullcirc & ~\cite{sun2022coprotector,ji2022unlearnable} \\
\cline{1-1}
\cline{9-9}
Watermark Defenses & \emptycirc & \fullcirc & \fullcirc & \fullcirc & \fullcirc & \emptycirc & \fullcirc & ~\cite{li2023protecting,guan2024codeip,li2024resilient,sun2023codemark} \\
\bottomrule
\end{tabularx}
\end{table*}

\textbf{Pre-processing defenses.} Pre-processing defenses mitigate the privacy risks of CLMs by filtering out the sensitive information at the training stage and blocking out the suspicious queries at the inference stage~\cite{yang2024unveiling,finkman2024codecloak}. Existing pre-processing-based defenses can be divided into dataset cleansing~\cite{yang2024unveiling} and input pre-processing~\cite{finkman2024codecloak}. The dataset cleansing~\cite{yang2024unveiling} removes sensitive personal information and duplicated data, to prevent CLMs from leaking these sensitive data. The input pre-processing for large CLMs~\cite{finkman2024codecloak} manipulates user prompts to prevent privacy leakage. 

\textbf{Post-processing defenses.} The post-processing defenses modify the model outputs to prevent privacy leakage~\cite{yang2024unveiling,yang2023gotcha}. For instance, Yang et al.~\cite{yang2024unveiling} appealed for post-hoc checks for memorization effect. Yang et al.~\cite{yang2023gotcha} proposed that switching the decoding strategy of large CLMs from beam search to top-k sampling can largely lower privacy leakage, such as membership leakage.

\textbf{Poisoning-as-a-defense.} Poisoning-as-a-defense leverages data poisoning to protect the user data privacy~\cite{ji2022unlearnable,sun2022coprotector}. The poisoning-as-a-defense injects perturbations into the private dataset such that the CLM accuracy will be damaged if the private dataset is illegally used for model training. However, this approach may be questionable because perturbed code examples transferred from a local surrogate model may conversely increase the generalization capability of the victim model if proper fine-tuning techniques are adopted~\cite{yang2024exploiting}. 

\textbf{Watermark defenses.} The watermark defenses protect the user data privacy by injecting copyright watermarks into the data and the model~\cite{sun2023codemark,sun2022coprotector,li2023protecting,guan2024codeip,li2024resilient}. Existing CLM watermark defenses against data extraction attacks can be divided into dataset watermarking~\cite{sun2023codemark,sun2022coprotector} and model watermarking~\cite{li2023protecting,guan2024codeip,li2024resilient}. The dataset watermarking~\cite{sun2022coprotector,sun2023codemark} aims to protect privacy by tracing the unauthorized usage of private data for CLM training. For instance, Coprotector~\cite{sun2022coprotector} leveraged targeted poisoning attacks as the dataset watermarking technique such that the CLM will output a pre-defined copyright message if trained on the poisoned dataset. CodeMark~\cite{sun2023codemark} proposed an imperceptible watermarking technique that is imperceptible to the human checker and the rule-based data filter. The model watermarking~\cite{li2023protecting,guan2024codeip,li2024resilient} is designed to protect the model copyright against model imitation attacks. For instance, LLWM~\cite{li2023protecting} embedded watermarks into the distribution of the output tokens of the target CLMs. CodeIP~\cite{guan2024codeip} embedded watermarks into the output logit distribution of the protected large CLMs. A grammar checker is applied such that the generated code can preserve the syntax. ACW method~\cite{li2024resilient} embedded the watermark into the model by applying the idempotent code transformations, such that the generated code can be traced back to its source CLM if the same code transformations can be correctly applied.

\section{Connections between each Risk} \label{sec:connections}
Each security aspect of CLMs is closely connected. The application of certain evasion attack/defense techniques may increase/decrease the risks from other aspects. However, there is little research in the code domain discussing the connections between different risks. To address this issue, we conducted a comparative literature review between the image and the code domain. Specifically, we first investigated the security connection studies in the image domain and then searched whether the code domain has similar studies. In this way, we can figure out the consistencies/inconsistencies of the research findings between the image and code domain, as well as point out the research gaps in the code domain. Our comparative review results are listed as Table~\ref{tab:risk_connections}, in which we summarize whether the application of certain attack/defense technology will lead to an increase (denoted as $\textcolor{red}{\boldsymbol{\uparrow}}$) / decrease (denoted as $\textcolor{green}{\boldsymbol{\downarrow}}$) / not sure (denoted as \textcolor{yellow}{\rotatebox[origin=c]{90}{\boldsymbol{$\leftrightarrows$}}}) / unexplored (denoted as $\textcolor{gray}{\boldsymbol{\cdots}}$) of risks from other aspects. In the following subsections, we will analyze each connection point in detail.

\begin{table*}[ht]
\scriptsize
\centering
\caption{The connections between each risk: comparing the image and the code domain. This table summarizes whether the application of certain attack/defense techniques will lead to an increase (denoted as \textcolor{red}{\rotatebox[origin=c]{180}{\boldsymbol{$\downarrow$}}}) / decrease (denoted as $\textcolor{green}{\boldsymbol{\downarrow}}$) / not sure (denoted as \textcolor{yellow}{\rotatebox[origin=c]{90}{\boldsymbol{$\leftrightarrows$}}}) / unexplored (denoted as $\textcolor{gray}{\boldsymbol{\cdots}}$) of risks from other aspects. The \textbackslash{} denotes the connection within the same risk, which is meaningless.}~\label{tab:risk_connections}
\begin{tabularx}{\textwidth}{>{\centering\arraybackslash}X>{\centering\arraybackslash}X|cccccc}
\toprule
\multicolumn{2}{c|}{\multirow{2}{*}{Attacks \& Defenses}} & \multicolumn{2}{c}{Evasion Risks} & \multicolumn{2}{c}{Poisoning Risks} & \multicolumn{2}{c}{Privacy Risks} \\
\multicolumn{2}{c|}{} & image & code & image & code & image & code \\
\midrule
\multicolumn{2}{c|}{Evasion Attack} & \textbackslash{} & \textbackslash{} & \textcolor{red}{\rotatebox[origin=c]{180}{\boldsymbol{$\downarrow$}}} ~\cite{shafahi2018poison} & $\textcolor{gray}{\boldsymbol{\cdots}}$ & \textcolor{red}{\rotatebox[origin=c]{180}{\boldsymbol{$\downarrow$}}} ~\cite{choquette2021label} & \textcolor{red}{\rotatebox[origin=c]{180}{\boldsymbol{$\downarrow$}}} ~\cite{zhang2023code,majdinasab2024trained} \\
\cline{1-2}
\multicolumn{1}{c|}{\multirow{2}{*}{\makecell{Evasion Defenses}}} & Adv. Training & $\textcolor{green}{\boldsymbol{\downarrow}}$~\cite{madry2017towards} & $\textcolor{green}{\boldsymbol{\downarrow}}$~\cite{jia2023clawsat,gao2023discrete,yefet2020adversarial,bielik2020adversarial,henkel2022semantic} & \textcolor{yellow}{\rotatebox[origin=c]{90}{\boldsymbol{$\leftrightarrows$}}}~\cite{tao2021better,wen2023adversarial}   & $\textcolor{gray}{\boldsymbol{\cdots}}$  & \textcolor{red}{\rotatebox[origin=c]{180}{\boldsymbol{$\downarrow$}}}~\cite{hayes2020trade,khaled2022careful,mejia2019robust}    & $\textcolor{gray}{\boldsymbol{\cdots}}$ \\
\cline{2-2}
\multicolumn{1}{c|}{} & Input Denoising & $\textcolor{green}{\boldsymbol{\downarrow}}$~\cite{grosse2017statistical} & $\textcolor{green}{\boldsymbol{\downarrow}}$~\cite{yefet2020adversarial,bielik2020adversarial,tian2023fly} & $\textcolor{green}{\boldsymbol{\downarrow}}$~\cite{tian2022comprehensive,mu2023progressive} & $\textcolor{green}{\boldsymbol{\downarrow}}$~\cite{hussain2023occlusion} & \textcolor{red}{\rotatebox[origin=c]{180}{\boldsymbol{$\downarrow$}}}~\cite{carlini2022privacy,duddu2021shapr,jayaraman2022attribute}    & $\textcolor{gray}{\boldsymbol{\cdots}}$ \\
\hline
\multicolumn{2}{c|}{Poisoning Attack} & \textcolor{red}{\rotatebox[origin=c]{180}{\boldsymbol{$\downarrow$}}}~\cite{tao2022can}   & $\textcolor{gray}{\boldsymbol{\cdots}}$  & \textbackslash{} & \textbackslash{} & \textcolor{red}{\rotatebox[origin=c]{180}{\boldsymbol{$\downarrow$}}}~\cite{chen2022amplifying,tramer2022truth} & $\textcolor{gray}{\boldsymbol{\cdots}}$ \\
\cline{1-2}
\multicolumn{1}{c|}{\multirow{2}{*}{\makecell{Poisoning Defenses}}} & Input Denoising    & $\textcolor{green}{\boldsymbol{\downarrow}}$~\cite{grosse2017statistical} & $\textcolor{green}{\boldsymbol{\downarrow}}$~\cite{yefet2020adversarial,bielik2020adversarial,tian2023fly} & $\textcolor{green}{\boldsymbol{\downarrow}}$~\cite{tian2022comprehensive,mu2023progressive} & $\textcolor{green}{\boldsymbol{\downarrow}}$~\cite{hussain2023occlusion} & \textcolor{red}{\rotatebox[origin=c]{180}{\boldsymbol{$\downarrow$}}}~\cite{carlini2022privacy,duddu2021shapr,jayaraman2022attribute}    & $\textcolor{gray}{\boldsymbol{\cdots}}$ \\
\cline{2-2}
\multicolumn{1}{c|}{} & Model Fine-tuning & $\textcolor{gray}{\boldsymbol{\cdots}}$ & $\textcolor{gray}{\boldsymbol{\cdots}}$  & $\textcolor{green}{\boldsymbol{\downarrow}}$~\cite{pang2023backdoor} & $\textcolor{green}{\boldsymbol{\downarrow}}$~\cite{liu2018fine,li2023multi} & $\textcolor{gray}{\boldsymbol{\cdots}}$ & $\textcolor{gray}{\boldsymbol{\cdots}}$ \\
\hline
\multicolumn{2}{c|}{Privacy Attack} & \textcolor{red}{\rotatebox[origin=c]{180}{\boldsymbol{$\downarrow$}}}~\cite{papernot2017practical}   & \textcolor{red}{\rotatebox[origin=c]{180}{\boldsymbol{$\downarrow$}}}~\cite{li2024extracting}   & $\textcolor{gray}{\boldsymbol{\cdots}}$   & $\textcolor{gray}{\boldsymbol{\cdots}}$   & \textbackslash{}   & \textbackslash{}    \\  
\cline{1-2}
\multicolumn{1}{c|}{\makecell{Privacy Defense}} & Watermarking & $\textcolor{gray}{\boldsymbol{\cdots}}$  & $\textcolor{gray}{\boldsymbol{\cdots}}$  & \textcolor{red}{\rotatebox[origin=c]{180}{\boldsymbol{$\downarrow$}}}~\cite{adi2018turning,li2022untargeted,zhang2018protecting}   & \textcolor{red}{\rotatebox[origin=c]{180}{\boldsymbol{$\downarrow$}}}~\cite{li2023protecting,guan2024codeip,li2024resilient}  & \textcolor{yellow}{\rotatebox[origin=c]{90}{\boldsymbol{$\leftrightarrows$}}}~\cite{lukas2022sok,tramer2022truth} & $\textcolor{gray}{\boldsymbol{\cdots}}$ \\
\bottomrule
\end{tabularx}
\end{table*}

\subsection{Evasion}

\textbf{Connections with poisoning risks.} 

(1) $\textcolor{red}{\boldsymbol{\uparrow}}$ Adversarial examples can improve the stealthiness of the poisoning attacks. In the image domain, adversarial examples can be leveraged to realize the ``clean-label'' poisoning attack~\cite{shafahi2018poison} to improve the attack's stealthiness. Assuming the poisoning adversary can not control the data labeling process of the victim system, the clean-label attack achieves targeted poisoning manipulation by adding adversarial perturbations from the source class into the target class. Currently, there is no research in the code domain focusing on how to leverage adversarial examples to enhance the stealthiness and effectiveness of the poisoning attacks, which deserves to be studied in future work.

(2) \textcolor{yellow}{\rotatebox[origin=c]{90}{\boldsymbol{$\leftrightarrows$}}} Adversarially trained models may still be vulnerable to poisoning attacks. Adversarial training is regarded as the most effective defense against adversarial examples~\cite{madry2017towards}. Researchers in the image domain wonder whether adversarial training can defend against both adversarial and poisoning examples. Tao et al.~\cite{tao2021better} stated that by suppressing the spurious features, adversarial training may serve as a universal defense for both adversarial and poisoning attacks. However, this claim is questioned by Wen et al.~\cite{wen2023adversarial} by crafting poisoning examples with stable features to undermine the effectiveness of adversarial training. Currently, there is no research in the code domain focusing on universal defense for both adversarial and poisoning attacks, which deserves to be studied in future work.

(3) $\textcolor{green}{\boldsymbol{\downarrow}}$ Input denoising defenses are both effective against adversarial and poisoning attacks. This conclusion is consistent in both image~\cite{tian2022comprehensive,jin2022can,mu2023progressive} and code domains~\cite{hussain2023occlusion,tian2023fly}. In the image domain, Jin et al.~\cite{jin2022can} leverage the commonsense of adversarial and poisoning perturbations to detect both adversarial and poisoning examples, such as outlier features, activated neurons, output scores, etc. Mu et al.~\cite{mu2023progressive} leveraged adversarial perturbations to enhance the effectiveness of backdoor detecting/erasing defenses. In the code domain, the outlier-detection-based defenses are also verified to be robust against both adversarial perturbations~\cite{tian2023fly} and poisoning triggers~\cite{hussain2023occlusion}.

\textbf{Connections with privacy risks.} 

(1) $\textcolor{red}{\boldsymbol{\uparrow}}$ Adversarial examples improve the effectiveness of black-box privacy attacks. For both the image and the code domain, adversarial examples can be utilized to enhance the effectiveness of privacy attacks. In both the image~\cite{choquette2021label} and the code domain~\cite{zhang2023code,majdinasab2024trained}, adversarial examples can be utilized to achieve label-only black-box membership inference attacks in terms of the different sensitivity of member/non-member data against adversarial examples. Despite the existing research progress, the connections between evasion attacks and other privacy attacks (e.g. data extraction attacks, model imitation attacks, etc) in the code domain still deserve exploration.

(2) \textcolor{yellow}{\rotatebox[origin=c]{90}{\boldsymbol{$\leftrightarrows$}}} Adversarial training increases the privacy risks. In the image domain, adversarially trained models are verified to be more vulnerable to membership inference attacks~\cite{hayes2020trade}, data extraction attacks~\cite{mejia2019robust}, and model imitation attacks~\cite{khaled2022careful} because of the smoothed decision boundary and the aggravated overfitting issue. Currently, there is no research in the CLM domain focusing on whether adversarial training will increase privacy risks, which deserves to be studied in future work.

(3) $\textcolor{red}{\boldsymbol{\uparrow}}$ Input denoising increases privacy risks. In the image domain, researchers have found that the input denoising defenses are vulnerable to membership inference attacks~\cite{carlini2022privacy} and data extraction attacks~\cite{jayaraman2022attribute}. This is because detecting outlier adversarial samples usually worsens the memorization effect, making the victim model more vulnerable to privacy leakage. Besides, researchers have identified the ``onion effect'' in input denoising defenses that when the layer of easiest-to-privacy-leakage examples is removed, another layer below becomes easy-to-privacy-leakage~\cite{carlini2022privacy}. Currently, there is no research in the code domain focusing on whether input denoising will increase the privacy leakage risks of the CLMs, which deserves to be studied in future work.

\subsection{Poisoning}

 \textbf{Connections with adversarial example risks.} 
 
$\textcolor{red}{\boldsymbol{\uparrow}}$ Poisoning attack increases the adversarial example risks. In the image domain, as mentioned above, researchers have found that adversarial training may not be robust against poisoning attacks~\cite{wen2023adversarial}. Based on this phenomenon, the poisoning adversary can target the adversarial training to destroy the robustness of the victim model~\cite{tao2022can}, making the victim model more vulnerable to evasion attacks at the test time. Currently, there is no research in the code domain focusing on the connections between poisoning attacks and evasion attacks, which deserves to be studied in future work.

\textbf{Connections with privacy risks.} 

$\textcolor{red}{\boldsymbol{\uparrow}}$ Poisoning attack increases the privacy risks. In the image domain, poisoned models will be more susceptible to privacy leakages. For instance, Chen et al.~\cite{chen2022amplifying} and Tramer et al.~\cite{tramer2022truth} illustrated that poisoned models are more vulnerable to membership inference attacks compared to normally trained counterparts. Currently, there is no research in the code domain focusing on the connections between poisoning attacks and privacy attacks, which deserves to be studied in future work.

\subsection{Privacy}

\textbf{Connections with adversarial example risks.} 

$\textcolor{red}{\boldsymbol{\uparrow}}$ Privacy attack increases the adversarial risks in the black-box setting. Privacy attacks can serve as preposition procedures for evasion attacks. This conclusion is consistent for both image and code domains. For instance, the model imitation attack extracts the functionality of the victim model thus enabling the success of transfer-based evasion attacks~\cite{papernot2017practical,li2024extracting}. For further work, the connections between privacy attacks and evasion attacks deserve to be further studied.

\textbf{Connections with poisoning risks.}

(1) $\textcolor{red}{\boldsymbol{\uparrow}}$ Watermarking is vulnerable to adaptive poisoning attacks. Watermarking protects the copyright of the model checkpoint by leveraging the poisoning scheme, making itself still vulnerable to adaptive poisoning attacks. This conclusion is consistent for both the image~\cite{adi2018turning,li2022untargeted,zhang2018protecting} and code~\cite{li2023protecting,guan2024codeip,li2024resilient} domains. For further work, watermark defenses robust to adaptive attacks deserve to be studied.

(2) \textcolor{yellow}{\rotatebox[origin=c]{90}{\boldsymbol{$\leftrightarrows$}}} Watermarking is robust to model imitation attacks while vulnerable to other privacy attacks. In the image domain, researchers have discovered that although watermarking can protect the model copyright, it is more vulnerable to membership inference attacks, data extraction attacks, and property inference attacks~\cite{lukas2022sok,chen2022amplifying,tramer2022truth}. The logic behind this phenomenon is that poisoned models are more susceptible to privacy leakage, as discovered by Chen et al.~\cite{chen2022amplifying} and Tramer et al.~\cite{tramer2022truth}. As a poisoning scheme, watermarking is thus also more vulnerable to the above privacy attacks than their standardly trained counterparts. Currently, there is no research in the code domain focusing on the connections between watermarking and other privacy inference attacks, which deserves to be studied in future work.

\section{Interpreting the Security Threats for CLMs: an XAI Point of View} \label{sec:XAI}

XAI can be a powerful tool for interpreting and analyzing CLMs, which can be divided into XAI tools from the model perspectives and the data perspectives~\cite{han2023interpreting}. This section first briefly reviews the XAI tools and then analyzes how these XAI tools interpret the poisoning, evasion, and privacy risks of CLMs. 


\subsection{XAI Tools for CLMs}

\textbf{XAI tools from the model perspectives.} The XAI tools from the model perspectives include the post-hoc feature attribution analysis~\cite{sotgiu2022explainability,steenhoek2023language,liu2024reliability}, the attention-based analysis, the neuron-based analysis, the probing-based analysis, and the causal analysis. The post-hoc feature attribution analysis~\cite{sotgiu2022explainability,steenhoek2023language,liu2024reliability} applies feature attribution-based XAI tools such as SHAP~\cite{lundberg2017unified} to interpret the behavior of CLMs~\cite{sotgiu2022explainability,liu2024reliability}. The attention-based interprets the mechanism of CLMs by analyzing the widely adopted attention module~\cite{vaswani2017attention} in CLMs~\cite{wan2022they,steenhoek2023language,saad2023naturalness}, which can be further divided into the attention-weights-based~\cite{wan2022they,steenhoek2023language} and the attention-distribution-based approaches~\cite{saad2023naturalness}. The neuron-based analysis interprets the CLMs from the perspective of hidden neurons~\cite{harel2020neuron,sharma2023interpreting} with the neuron coverage~\cite{harel2020neuron} and the neuron reduction~\cite{sharma2023interpreting}. The probing-based analysis designs probing tasks to interpret CLMs~\cite{karmakar2021pre,wan2022they,troshin2022probing,ma2022code,hernandez2022ast,shi2023towards,karmakar2023inspect}, such as the surface-level probing tasks~\cite{karmakar2021pre,karmakar2023inspect}, syntactic-level probing tasks\cite{karmakar2021pre,wan2022they,troshin2022probing,hernandez2022ast,karmakar2023inspect}, structure-level probing tasks~\cite{karmakar2021pre,karmakar2023inspect}, and semantic-level probing tasks~\cite{karmakar2021pre,troshin2022probing,ma2022code}. Causal analysis~\cite{rodriguez2023benchmarking,palacio2024toward} eliminate spurious correlations between the treatments $T$ and the outcomes $Y$ by controlling confounding features $Z$. 

\textbf{XAI tools from the data perspectives.} XAI tools from the data perspective can be categorized into the rule induction~\cite{cito2021explaining}, the data distribution analysis~\cite{hajipour2022simscood}, and the memorization effect~\cite{rabin2023memorization}. The rule induction~\cite{cito2021explaining} divides wrongly predicted data in terms of certain statistical criteria as an interpretation for CLMs. The data distribution analysis~\cite{hajipour2022simscood} interpreted CLMs from the out-of-distribution (OOD) generalization perspective. The memorization effect~\cite{rabin2023memorization} points out that the large model capacities of CLMs tend to make CLMs memorize the training data, leading to sever security and privacy issues.

\subsection{Interpreting CLM Risks using XAI Tools}

\begin{table*}[ht]
\scriptsize
\caption{Summary of XAI insights for adversarial machine learning on CLMs.}~\label{tab:summary_of_XAI_insights}
\centering
\begin{tabularx}{\textwidth}{>{\cellcolor{white}\centering}m{1.5cm}|>{\centering\arraybackslash}m{2.0cm}>{\centering\arraybackslash}X>{\centering\arraybackslash}X>{\centering\arraybackslash}X}
\toprule
\multirow{1}{*}{Perspective}                        & \multirow{1}{*}{XAI Method} & XAI Insights for Poisoning   Attacks    & XAI Insights for Evasion Attacks      & XAI Insights for Privacy Attacks     \\
\midrule
\rowcolor{gray!20}  &  \multirow{4}{*}{{\centering \makecell{Post-hoc \\ feature \\ attribution \\ analysis}}} & \multirow{3}{*}{\centering \textbackslash{}}             
& ~\cite{liu2024reliability,palacio2024toward}:  CLMs are sensitive to semantic-preserving transformations used in adversarial examples.   & \multirow{4}{*}{\centering \textbackslash{}}   \\

&  \multirow{4}{*}{\centering \makecell{Attention-based \\ analysis}}   & ~\cite{hussain2024measuring_new}:  Poisoning triggers have little impact on attention weights, making it hard to detect backdoors on CLMs.              & ~\cite{zhang2022diet,troshin2022probing,saad2023naturalness}:  CLMs are sensitive to semantic-preserving transformations used in adversarial examples. & \multirow{4}{*}{\centering \textbackslash{}}      \\

\rowcolor{gray!20}  &  \multirow{4}{*}{\centering \makecell{Neuron-based \\ analysis}}        & ~\cite{hussain2024measuring_new}:  Poisoning triggers have little impact on neuron activations, making it hard to detect backdoors on CLMs.             & ~\cite{harel2020neuron,sharma2023interpreting,wang2023distxplore}:  The generation of adversarial examples needs a more effective guidance metric other than neuron coverage.    & \multirow{4}{*}{\centering \textbackslash{}}   \\

& \multirow{3}{*}{\centering \makecell{Probing-based \\ analysis}}      & \multirow{3}{*}{\centering \textbackslash{}}                  & ~\cite{ahmed2023towards}: CLMs are robust to semantic-preserving transformations (contrary to the mainstream)      & \multirow{4}{*}{\centering \textbackslash{}}         \\

\rowcolor{gray!20} \multirow{-12}{*}{\makecell{\cellcolor{white} Model \\ \cellcolor{white} Perspective}} & \multirow{3}{*}{{\centering \makecell{Causal \\ analysis}}}   & \multirow{3}{*}{\centering \textbackslash{}}  & ~\cite{palacio2024toward}: CLMs are sensitive to semantic-preserving transformations used in adversarial examples. & \multirow{4}{*}{\centering \textbackslash{}}   \\

\midrule
&  \multirow{3}{*}{\centering Rule induction}   & \multirow{3}{*}{\centering \textbackslash{}}    & ~\cite{rabin2021understanding}:  CLMs can make decisions relying on non-important tokens, explaining why adversarial examples exist.    & \multirow{4}{*}{\centering \textbackslash{}}    \\

\rowcolor{gray!20} & \multirow{4}{*}{{\centering \makecell{Data \\ distribution \\ analysis}}}      & ~\cite{hussain2024measuring_new}:  Poisoned samples behave distinct from normal samples on context embeddings,  enabling the CLM poisoning detection. & \multirow{5}{*}{\centering \textbackslash{}}  &  \multirow{5}{*}{\centering \textbackslash{}}   \\

\multirow{-6}{*}{\makecell{\cellcolor{white} Data \\ \cellcolor{white} Perspective}}  & \multirow{3}{*}{{\centering \makecell{Memorization \\ effect}}}    & ~\cite{yang2024unveiling}: The memorization effect explains why poisoning a small fraction of data is enough against CLM.       & \multirow{4}{*}{\centering \textbackslash{}} &~\cite{al2024traces,yang2024unveiling,karmakar2022codex,carlini2022quantifying}:  The memorization effect is the root cause of the membership inference and the data extraction attack. \\
\bottomrule
\end{tabularx}
\end{table*}

The interpretations for each security risk are summarized as Table ~\ref{tab:summary_of_XAI_insights}, which will be analyzed in detail below.

\textbf{XAI insights for the poisoning attacks.}

(1) CLMs are prone to memorize a few poisoned samples due to their larger number of parameters, making it possible to manipulate the victim CLM with only a small fraction of poisoned samples. Yang et al.~\cite{yang2024unveiling} observed that larger models have a stronger memorization ability than smaller ones. The memorization effect enables the adversaries to keep low poisoning rates~\cite{yang2024stealthy,li2023poison}. Specifically, Aghakhani et al.~\cite{aghakhani2024trojanpuzzle} found that there was no significant decrease in the attack performance with a lower poisoning rate.

(2) The context embedding can be an effective metric for detecting poisoned samples. As has been verified by Ramakrishnan et al.~\cite{ramakrishnan2022backdoors} and Hussain et al.~\cite{hussain2024measuring_new}, input embeddings, neuron activations, and attention weights cannot distinguish between poisoned samples and normal samples because of the large amount of parameter space of CLMs. Alternatively, context embeddings~\cite{kanade2020learning} are regarded as promising metrics for detecting poisoning samples for CLMs~\cite{hussain2024measuring_new}, which are observed to be clustered into different groups for different trigger patterns.

\textbf{XAI insights for the evasion attacks.} 

(1) CLMs can make predictions relying on only a small fraction of syntax tokens, which serve as the fundamental explanation for the adversarial example vulnerability of CLMs~\cite{rabin2021understanding}. Rabin et al.~\cite{rabin2021understanding} interpreted the CLMs from the data perspective. They utilized the rule induction technique to analyze the influence of each token on the output of CLMs and found that CLMs can make predictions relying on only a small fraction of syntax tokens. This observation can explain why the adversary can mislead the prediction of victim CLMs with only minor changes. 

(2) Evasion attacks against CLMs need more effective attack guidance other than the traditional neuron coverage metric~\cite{harel2020neuron}. The neuron coverage metric is commonly adopted as the guidance of the adversarial example generation process against computer vision models~\cite{xie2019coverage,xie2022npc}, and the neuron coverage rate is positively correlated with the evasion attack success rate. However, this conclusion may not hold for CLMs where higher neuron coverage does not necessarily lead to higher attack success rates~\cite{harel2020neuron}. This finding calls for novel and more effective metrics to guide the adversarial example generation process against CLMs. As indicated by Wang et al.~\cite{wang2023distxplore} and Sharma~\cite{sharma2023interpreting} et al., neuron distribution can be a promising metric for the adversarial example generation.

(3) Middle-level features of CLMs can be leveraged to craft transferable adversarial examples and to enhance the generalization and robustness of CLMs~\cite{hernandez2022ast,ma2022code,shi2023towards,yang2024exploiting}. Hernandez et al.~\cite{hernandez2022ast}, Ma et al~\cite{ma2022code} and Shi et al.~\cite{shi2023towards} analyzed the embedded information across different layers of CLMs with the probing-based technique. They found that the lexical and syntax features embedded in the middle-level features can serve as a universal representation across different code snippets. Thus, leveraging the middle-level features of surrogate CLMs can craft highly transferable adversarial examples~\cite{yang2024exploiting}. Besides, the middle-level adversarial examples are also beneficial for adversarial fine-tuning~\cite{yang2024exploiting}. 
  
\textbf{XAI insights for the privacy attacks.} 

(1) The memorization effect can be one of the major causes of privacy issues of CLMs~\cite{karmakar2022codex,carlini2022quantifying,yang2024unveiling,al2024traces}, 
especially for the membership inference attack and the data extraction attack. On the one hand, the memorization effect makes CLMs behave differently on member and non-member data and thus can be exploited for membership inference~\cite{karmakar2022codex,al2024traces}. On the other hand, the memorization effect enables the adversary to extract the original training data by providing certain input prefixes~\cite{carlini2022quantifying,karmakar2022codex,yang2024unveiling,al2024traces}.

\section{Future Gazing} \label{sec:future}
Despite existing research progress, there are still many valuable research questions and research opportunities for further work, which we list below.

\textbf{Reliable defenses against poisoning attacks in realistic settings.} Future research opportunities for reliable defenses against poisoning attacks in realistic settings lie in the following two aspects. First, existing data poisoning defense methods for CLMs are easily circumvented by newly developed adaptive attacks, which is one of the major challenges and opportunities for reliable defenses against poisoning attacks on CLMs. Practically, strictly defining and systematizing the patterns of existing poisoning attacks are of vital importance for reliable defenses. Second, existing defenses commonly assume that the defenders have access to the original clean dataset, which is not practical because of privacy and property issues of the training data. It is valuable to develop poisoning defenses for CLMs under restricted defenders' knowledge,  where the defenders have no access to the original data.

\textbf{Cross-domain transfer-based evasion attacks against realistic CLM applications.} The cross-domain transfer-based evasion attacks against realistic CLMs can be one of the major research opportunities for CLMs. As pointed out by the latest research on practitioners~\cite{grosse2023towards}, the cross-domain transfer-based threat model is the most practical evasion attack threat model since the i.i.d. training data, the query score feedback, and the number of quires are commonly restricted for adversaries in the real world. However, the number of existing research on cross-domain transfer-based evasion attacks on CLMs is limited because of the discrete nature of source code.

\textbf{Privacy attacks in realistic black-box settings, especially against large CLMs.} The research opportunities for privacy attacks against CLMs in realistic settings lie in the following two aspects. First, studying more realistic privacy attack threat models. Existing privacy attack techniques are often limited in attacking real-world CLMs because of the limited number of adversarial queries. Instead, recovering privacy information from normal user inputs may be a more practical technique against realistic CLM systems~\cite{grosse2023towards}. Second, conducting privacy attacks against the latest commercial large CLMs. Privacy attacks against the latest commercial large CLMs including GPT-4 and Genimi can be highly challenging and valuable because they are ``real black-box'' systems.

\textbf{Connections between each risk on CLMs.} Studying the connections between each risk on CLMs can be an opportunity for future research. Existing research has built a comprehensive theoretical framework for investigating the connections between each risk in computer vision models, but there is little research discussing the connections between risks on CLMs. Understanding the inter-relationships of different aspects of security risks of CLMs is significant for building unified defense methods against all security risks, or at least, increasing the robustness of one aspect without sacrificing the robustness of other aspects.

\textbf{XAI studies focused on the risks of CLMs.} Interpreting the risks of CLMs with the XAI tools can also be an opportunity for future research. Existing research has paid much attention to the security and privacy risks of CLMs. However, there is a lack of comprehensive investigation of the intrinsic mechanism of the security and privacy risks of CLMs from the XAI point of view. Explaining the risks of CLMs is valuable for deeply understanding the limitations of CLMs as well as finding out ways to fix these risks.

\textbf{Security risks in the latest auto-regressive large CLMs.} Studying the security risks of the latest auto-regressive large CLMs including GPT-4, Genimi, and CodePilot can be one of the major opportunities for future research because of the applicability and popularity of large CLMs for solving real-world tasks. The different technical characteristics of the latest auto-regressive large CLMs compared to traditional CLMs have led to tremendous new security risks, including large CLM memory poisoning attacks~\cite{deng2024pandora,chen2024agentpoison}, jailbreak attacks~\cite{chao2023jailbreaking}, DoS evasion attacks~\cite{gao2024inducing}, prompt injection attacks~\cite{greshake2023not,liu2023prompt}, prompt leakage issues~\cite{agarwal2024investigating}, etc.

\section{Conclusion} \label{sec:conclusion}
This paper surveys the security risks of Code Language Models (CLMs) from the perspective of confidentiality, integrity, and availability, covering topics of poisoning attacks. evasion attacks. and privacy attacks. We collected 79 papers from the latest research in the fields of artificial intelligence, computer security, and software engineering. For each type of adversarial risk, we analyzed the threat models, attack techniques, and the corresponding defense methods. To gain a deeper understanding of the adversarial machine learning of CLMs, we adopted novel investigation perspectives of XAI and connections between each risk. Finally, we pointed out existing research challenges in the field of adversarial machine learning of CLMs that deserve to be discussed in future works. Our paper helps CLM researchers and practitioners with the current progress as well as provides future opportunities for developing more reliable and trustworthy CLMs.

{
\bibliographystyle{IEEEtran}
\bibliography{reference}
}

\begin{IEEEbiography}[{\includegraphics[width=1in,height=1.25in,clip,keepaspectratio]{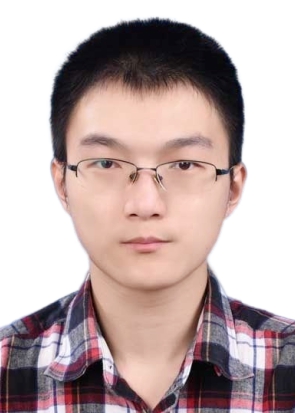}}]{Yulong Yang} received the B.S. degree in Computer Science and Technology from Xi'an Jiaotong University in 2022. He is currently pursuing a Ph.D. degree at the School of Cyber Science and Engineering, Xi'an Jiaotong University. His current research interests include adversarial machine learning, model compression \& acceleration, and multi-modal large models.
\end{IEEEbiography}

\begin{IEEEbiography}
[{\includegraphics[width=1in,height=1.25in,clip,keepaspectratio]{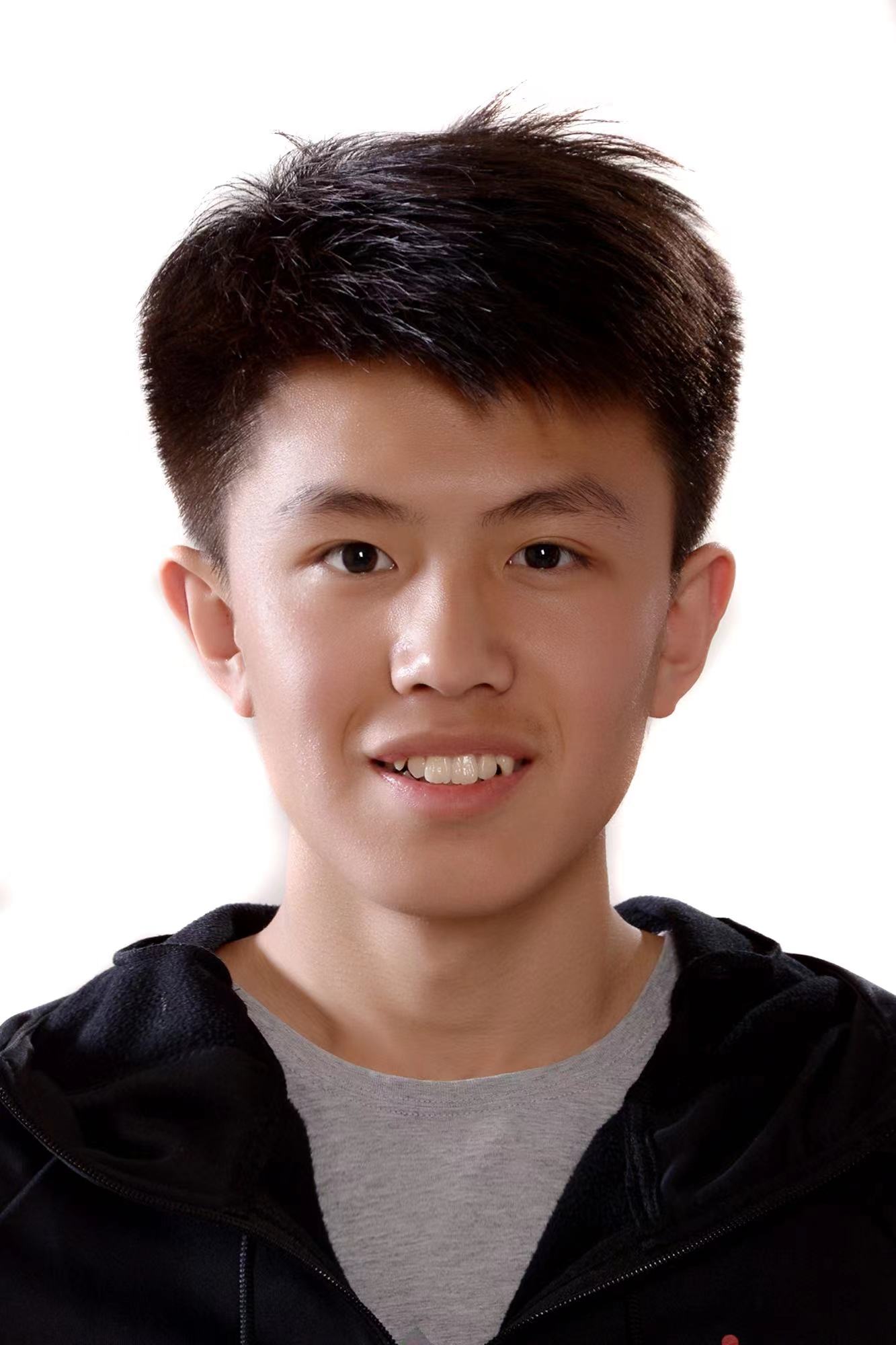}}]{Haoran Fan}
received the B.Eng. degree in Software Engineering from Dalian University of Technology in 2023 and is currently pursuing the M.Eng. degree in Software Engineering with the School of Software Engineering, Xi'an Jiaotong University. His current research interest is adversarial machine learning.
\end{IEEEbiography}

\begin{IEEEbiography}
[{\includegraphics[width=1in,height=1.25in,clip,keepaspectratio]{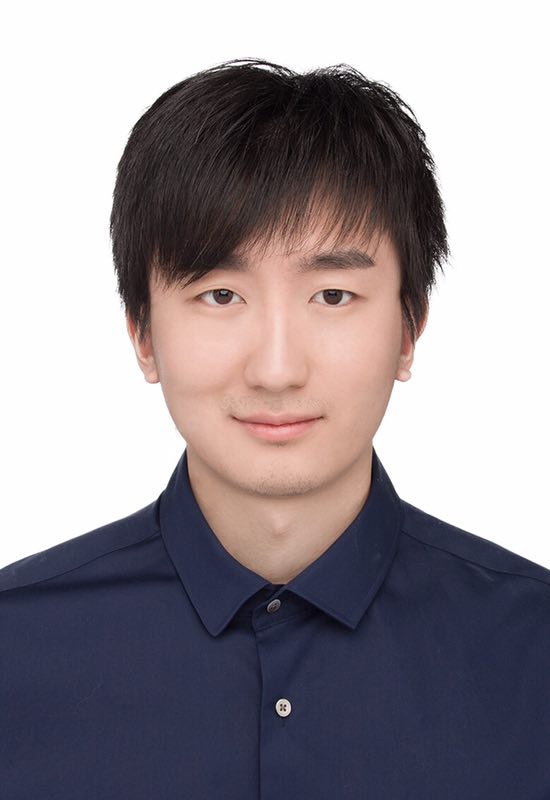}}]{Chenhao Lin} (Member, IEEE) received the B.Eng. degree in automation from Xi’an Jiaotong University in
2011, the M.Sc. degree in electrical engineering from Columbia University, in 2013, and the Ph.D. degree from The Hong Kong Polytechnic University, in 2018. He is currently a Research Fellow at the Xi’an Jiaotong University of China. His
research interests are in artificial intelligence security, identity authentication, biometrics, adversarial attack and robustness, and pattern recognition.
\end{IEEEbiography}

\begin{IEEEbiography}
[{\includegraphics[width=1in,height=1.25in,clip,keepaspectratio]{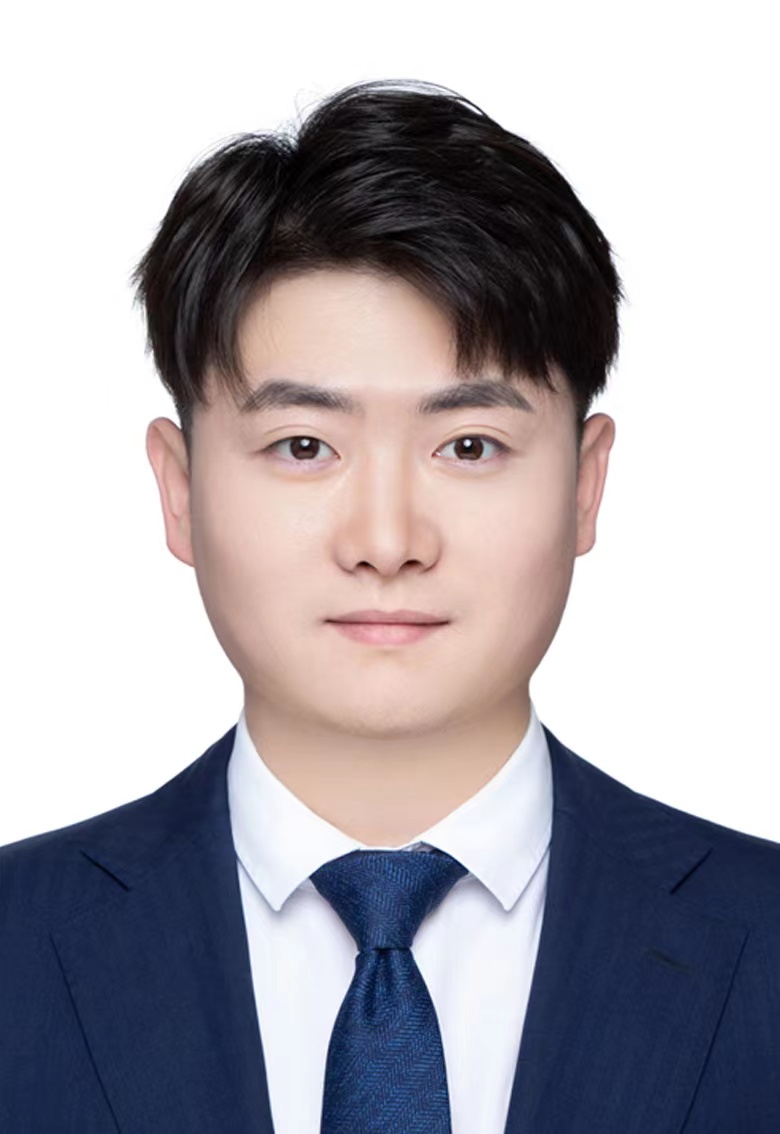}}] {Qian Li} (Member, IEEE) received the Ph.D. degree in computer science and technology from Xi’an Jiaotong University, China, in 2021. He is currently an Assistant Professor with the School of Cyber Science and Engineering, Xi’an Jiaotong University. His research interests include adversarial deep learning, artificial
intelligence security, and optimization of theory.
\end{IEEEbiography}

\begin{IEEEbiography}
[{\includegraphics[width=1in,height=1.25in,clip,keepaspectratio]{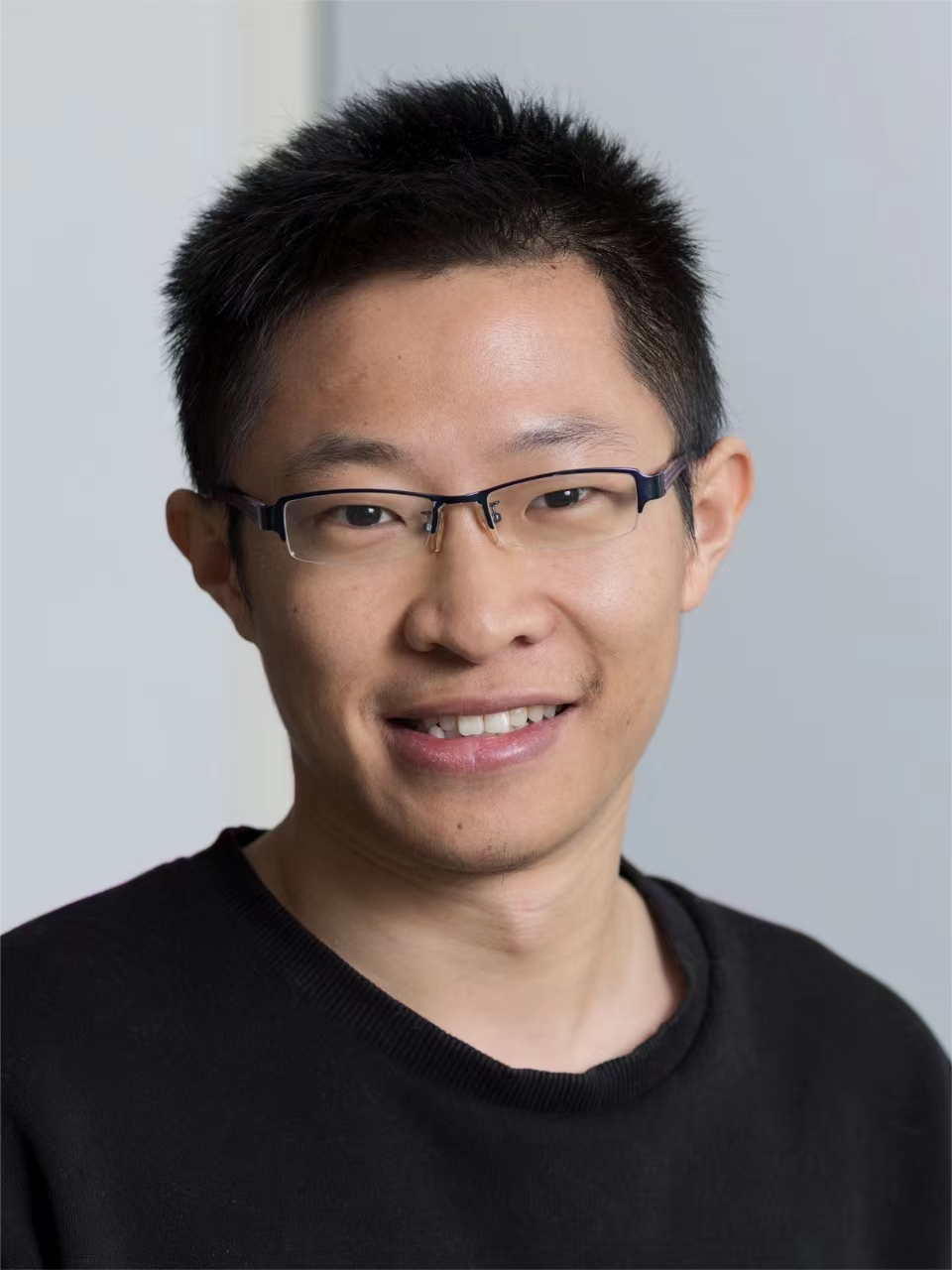}}]{Zhengyu Zhao} (Member, IEEE) received the Ph.D. degree from Radboud University, The Netherlands. He is currently an Associate Professor at Xi’an Jiaotong University, China. His general research interests include machine learning security and privacy. Most of his work has concentrated on security (e.g., adversarial examples and data poisoning) and privacy (e.g., membership inference) attacks against deep learning-based computer vision systems.
\end{IEEEbiography}

\begin{IEEEbiography}
[{\includegraphics[width=1in,height=1.25in,clip,keepaspectratio]{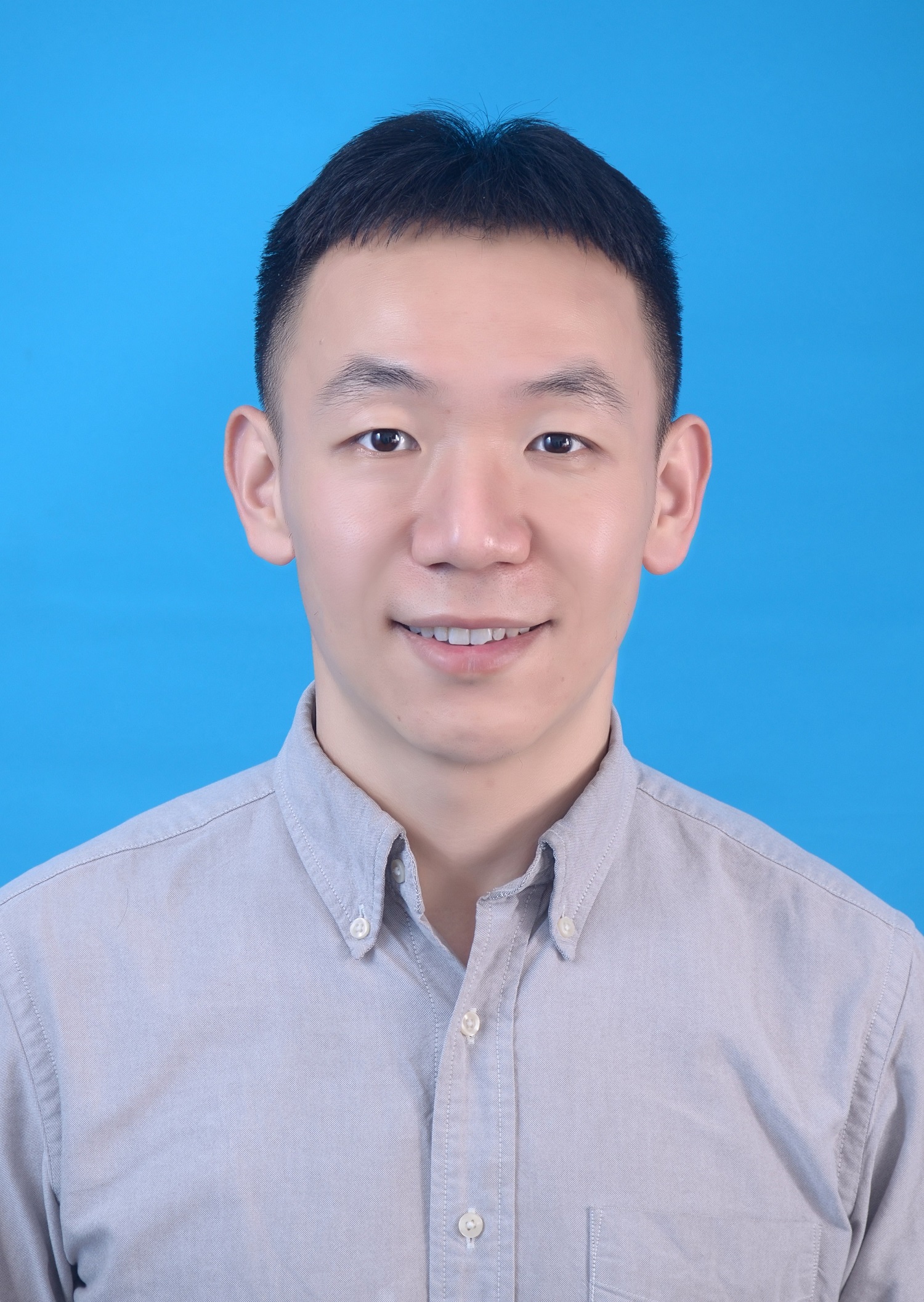}}]{Chao Shen} (Senior Member, IEEE)
received the B.S. degree in Automation from Xi'an Jiaotong University, China in 2007; and the Ph.D. degree in Control Theory and Control Engineering from Xi'an Jiaotong University, China in 2014. He is currently a Professor in the Faculty of Electronic and Information Engineering at Xi'an Jiaotong University. His current research interests include AI Security, insider/intrusion detection, behavioral biometrics, and measurement/experimental
methodology.
\end{IEEEbiography}

\begin{IEEEbiography}
[{\includegraphics[width=1in,height=1.25in,clip,keepaspectratio]{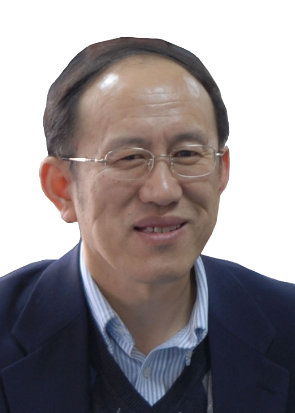}}]{Xiaohong Guan} (Fellow, IEEE) received the BS and MS degrees in automatic control from Tsinghua University, in 1982 and 1985, respectively, and the PhD degree in electrical engineering from the University of Connecticut, in 1993.  Since 1995, he has been with the Department of Automation, Tsinghua National Laboratory for Information Science and Technology, and the Center for Intelligent and Networked Systems, Tsinghua University. He is currently with the MOE Key Laboratory for Intelligent Networks and Network Security, Faculty of Electronic and Information Engineering, Xi'an Jiaotong University,
Xi'an, China, where he is also the dean of the Faculty
of Electronic and Information Engineering. He is an academician of the Chinese Academy of Sciences.
\end{IEEEbiography}

\end{document}